\documentclass[a4paper,11pt]{article}
\pdfoutput=1

\usepackage{jheppub}

\usepackage[T1]{fontenc}
\usepackage{xcolor}
\usepackage{amssymb}
\usepackage{amsmath}
\usepackage{graphicx}
\usepackage{subfig}
\usepackage{jheppub}
\usepackage{enumerate}
\usepackage{multirow}

\newwrite\todofile
\immediate\openout\todofile=\jobname.tdo

\newcounter{todocounter}

\newcommand{\printtodos}{
        \section*{To-Do List}
        \immediate\closeout\todofile
        \input{\jobname.tdo}
}

\newcommand{\la}[1]{\label{#1}}
\newcommand{\eq}[1]{\eqref{#1}}

\newcommand{\comment}[1]{}

\def\[{\left[}
\def\]{\right]}
\def\({\left(}
\def\){\right)}
\def\d{\partial}
\newcommand{\beq}{\begin{equation}}
\newcommand{\eeq}{\end{equation}}
\newcommand\beqa{\begin{eqnarray}}
\newcommand\eeqa{\end{eqnarray}}

\definecolor{color1}{rgb}{0.37800259999999997, 0.7777646, 0.4612972}
\definecolor{color2}{rgb}{0.0949579, 0.4544346, 0.5405524}
\definecolor{color3}{rgb}{1, 0.7508478000000001, 0.10824136000000001}
\definecolor{color4}{rgb}{0.9939926, 0.5927593999999999, 0.07059433999999999}
\definecolor{color5}{rgb}{0.969963, 0.376081, 0.0322881}
\definecolor{color6}{rgb}{0.7514516, 0.09778, 0.006329439999999999}

\title{\boldmath Computing Four-Point Functions with Integrability, Bootstrap and Parity Symmetry}

\author{Andrea Cavagli\`a$^a$}
\author{Nikolay Gromov$^{b}$}
\author{Michelangelo Preti$^{a,c,d}$}
\affiliation{
  $^a$ Dipartimento di Fisica, Universit\`a di Torino and INFN - Sezione di Torino\\
  Via P. Giuria 1, 10125 Torino, Italy
 \\
 $^b$ Department of Mathematics, King's College London\\
 Strand WC2R 2LS, London, UK \\
 $^c$ C. N. Yang Institute for Theoretical Physics, Stony Brook University\\
 Stony Brook, New York 11794, USA\\
 $^d$ Simons Center for Geometry and Physics, Stony Brook University\\
Stony Brook, New York 11794, USA
}
 \emailAdd{andrea.cavaglia@unito.it}
 \emailAdd{nikolay.gromov@kcl.ac.uk}
 \emailAdd{michelangelo.preti@stonybrook.edu}

\abstract{
The combination of integrability and crossing symmetry has proven to give tight non-perturbative bounds on some planar structure constants
in $\mathcal{N}$=4 SYM, particularly in the setup of defect observables built on a Wilson-Maldacena line. 
Whereas the precision is good for the low lying states, higher in the spectrum it drops due to the degeneracies at weak coupling when considering a single correlator. As this could be a clear obstacle in restoring higher point functions, we studied the problem of bounding directly a 4-point function at generic cross ratio, showing how to adapt for this purpose the numerical bootstrap  algorithms based on semidefinite programming. 
Another tool we are using to further narrow the bounds is a parity symmetry descending from the $\mathcal{N}$=4 SYM theory, which allowed us to reduce the number of parameters. We also give an interpretation for the parity in terms of the Quantum Spectral Curve at weak coupling. Our numerical bounds give an accurate determination of the 4-point function for physical values of the cross ratio, with at worst 5-6 digits precision at weak coupling and reaching more than 11 digits  for 't Hooft coupling $\frac{\sqrt{\lambda}}{4 \pi} \sim 4$. 
}

\begin{document} 
\maketitle
\flushbottom

\section{Introduction}
The integrability in theories like $\mathcal{N}$=4 SYM currently allows us to compute a number of observables, mostly in the planar limit.
This is the result of several years of developments in integrability in AdS/CFT correspondence, which are  ongoing, see e.g. 
\cite{Beisert:2010jr,Gromov:2017blm,Dorey:2019tcm} for reviews. In particular, the planar spectrum of local operators is under perfect control via the Quantum Spectral Curve (QSC) method~\cite{Gromov:2013pga,Gromov:2014caa}\footnote{While we are concerned with $\mathcal{N}$=4 SYM here, this method was also 
developed for ABJM theory~\cite{Cavaglia:2014exa,Bombardelli:2017vhk} and a QSC was also recently put forward for AdS$_3$/CFT$_2$ in the case of pure Ramond-Ramond flux~\cite{Ekhammar:2021pys,Cavaglia:2021eqr}. 
}. Numerical algorithms are 
available to obtain dimensions of short operators, including the recent C++ implementation which was used to generate the precise values of the dimensions of the first $219$ operators in a wide range of coupling~\cite{Gromov:2015wca,Hegedus:2016eop,Gromov:2023hzc}. At the leading order in the planar limit, the multi-point correlators of local operators factorise into 2-point functions, so in this sense the leading dynamics of local operators is under control.\footnote{Integrability should also know about multi-point correlators at $1/N$ orders, where they are nontrivial, for a continuous progress in integrable description of them see \cite{Basso:2015zoa,Bargheer:2017nne,Basso:2022nny} for developments in this direction.} Moving on, the gauge theories also have non-local observables such as light-ray operators (see \cite{Klabbers:2023zdz} for a recent QSC approach) and Wilson-lines, which are nontrivial already at leading order. Thus, understanding them is crucial for obtaining a complete solution of $\mathcal{N}$=4 SYM in the planar limit. 

Luckily, some non-local observables can be readily studied with QSC methods, such as two-point functions of operators inserted at the two cusps connected with circle segments of super-symmetric version of Wilson line~\cite{Gromov:2015dfa}. 
In principle, such a Wilson line of any shape can be re-expanded around the straight line with the so-called contour deformations (or tilt) 
operators~\cite{Cooke:2017qgm}.
Those operators are protected and the case of two insertions is well studied both with localisation and integrability~\cite{Correa:2012nk,Correa:2012at,Drukker:2012de, Gromov:2013qga}.  The  four-point function of these insertions is a highly nontrivial object, for which several perturbative orders at weak and strong coupling were computed with various techniques
analytically~\cite{Kiryu:2018phb,Cavaglia:2022qpg,miscPeveriBarrat,Giombi:2017cqn,Ferrero:2021bsb,Ferrerotoaappear}, see also
\cite{Barrat:2022eim,Barrat:2021tpn,Giombi:2023zte} for related further results on the theory. 

\begin{figure}
    \centering
    \includegraphics[width=\dimexpr 0.49\columnwidth\relax]{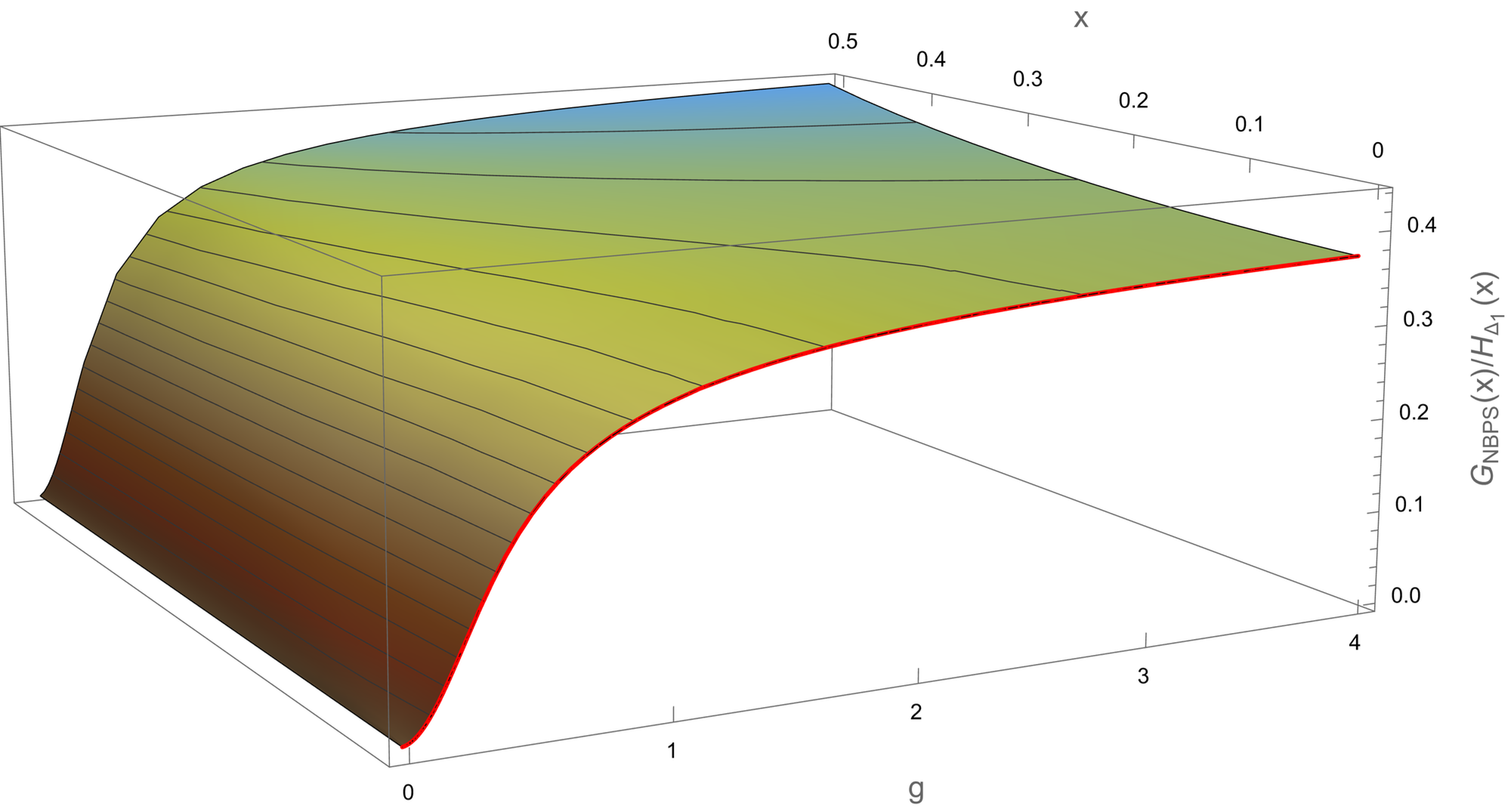}
    \includegraphics[width=\dimexpr 0.49\columnwidth\relax]{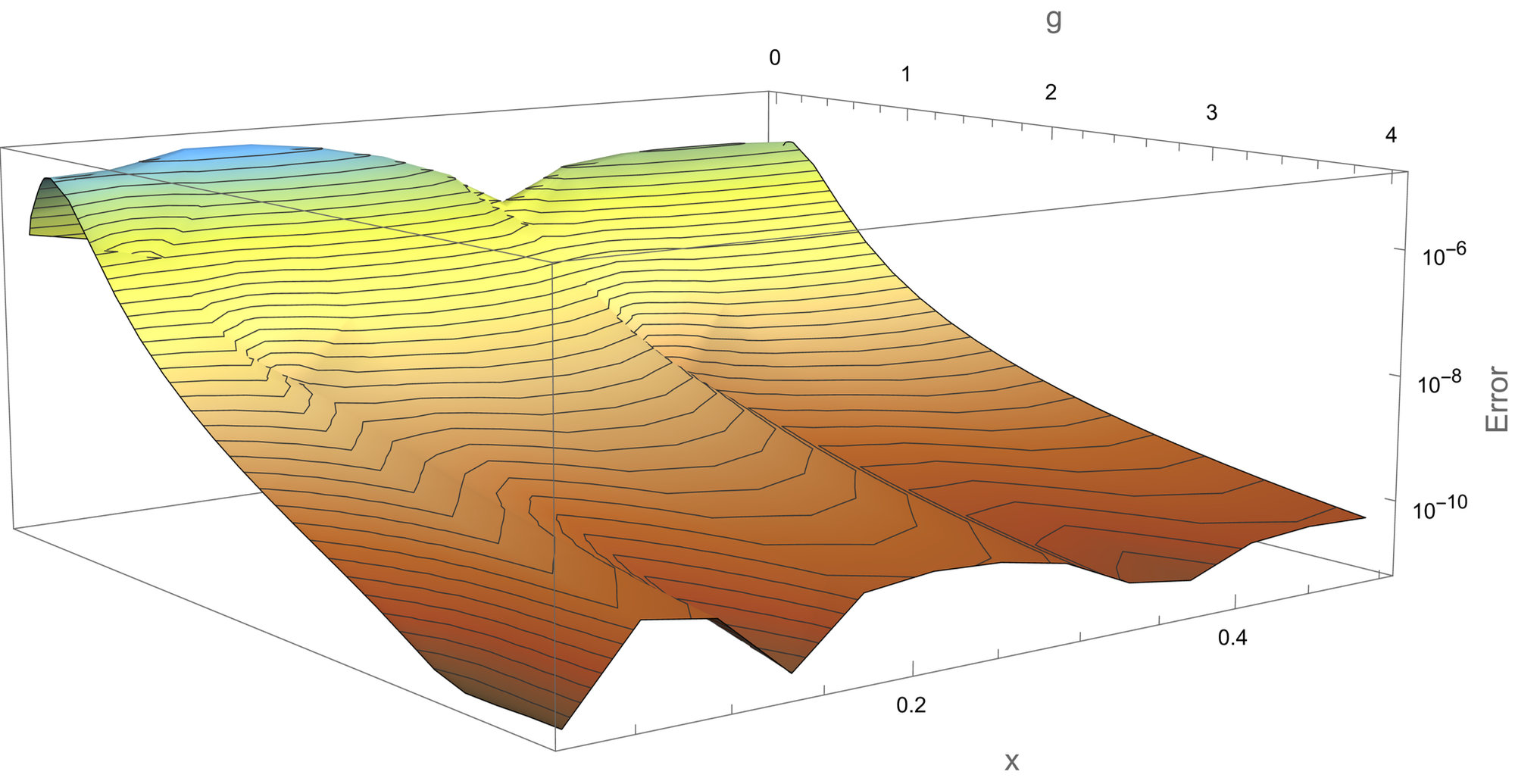}
    \caption{
    In the left plot we show the bound for the non-BPS part of the 4-point function amplitude $G(x)$, divided by the function $H_{\Delta_1}$ defined in (\ref{eq:chooseHDelta2})  
    (we refer to notations explained in the paper). The plot shows actual lower and upper bounds, which are so close that they give the impression of an exact plot. The normalisation by $H_{\Delta_1}$ is chosen in such a way that for for cross ratio $x=0$ the function should coincide with the leading OPE coefficient $C_1^2$. Indeed, the red line agrees with the bounds for this quantity found in our previous paper \cite{Cavaglia:2021bnz} with Julius Julius. As the plot illustrates,  in this paper an extra parameter is switched on, giving the dependence on the cross ratio $x$ of the 4-point function.
    The right plot represent the error in Log scale defined as $(G_{\text{upper}}-G_{\text{lower}})/2$.
    For a more detailed discussion of the results (and more recognisable plots of the 4-point function) see section \ref{sec:4pt}.   
   }
    \label{fig:intro}
\end{figure}

Integrability allows to compute conformal dimensions of the operators appearing in the OPE expansion of two contour-deforming operators. Using this information and the constraints due to the crossing symmetry, in \cite{Cavaglia:2022yvv,Cavaglia:2022qpg} very accurate numerical values were obtained for the OPE coefficients of some operators (see also \cite{Niarchos:2023lot} for a machine learning approach). In principle one can use those values to reconstruct the full 4-point function -- however, the precision would be hard to control due to a limited precision on the bounds on the structure constants for highly excited states situated in the denser part of the spectrum. 

However this impasse can be avoided by asking a sharp question directly on the value of the 4-point function. In this paper we obtain bounds directly for the value of the 4-point function at generic cross ratio, adapting for this purpose the numerical conformal bootstrap \texttt{SDPB} algorithm (see \cite{Poland:2018epd,Chester:2021aun,Poland:2022qrs,Rychkov:2023wsd} for reviews). Recently, bounds for the value of 4-point functions in CFTs were studied with the conformal bootstrap both analytically and numerically in \cite{Paulos:2020zxx,Paulos:2021jxx,Antunes:2021abs}\footnote{and earlier, in \cite{Lin:2015wcg},  a lower bound was obtained for a 4-point function in the $K_3$ CFT$_2$ using an integrated correlator identity.}, and we will show how our theory sits within those bounds. 
Our result is illustrated in figure \ref{fig:intro}, and in more detail in section \ref{sec:4pt}. 
Thanks to the information coming from integrability, in our case we can zoom on our individual theory and obtain a remarkably accurate plot of the 4-point function with several digits of precision at any coupling.

In order to further tighten the precision, we use a previously underexploited parity symmetry\footnote{We thank Ant\'onio Antunes for discussing his complementary results at strong coupling.} of the defect CFT, detailing  its QFT origin as well as explaining how to compute the corresponding $\mathbb{Z}_2$ charges of operators from their  integrability description. We explain how, as a result of this discrete symmetry, an 
infinite subset of all structure constants vanishes exactly for any value of the coupling constant. We also explain how the parity symmetry imposes reality constraints on another infinite class of structure constants, reducing the amount of the CFT data entering into the bootstrap procedure, and which becomes relevant especially for mixed correlators setups~\cite{MultiC}.

The paper is organized as follows. Section \ref{sec:setup} lays out the setup and main notations for the paper. In Section \ref{sec:parity_symmetry}, we explore the parity symmetry of 1D defect theory. The derivation of the parity operator within an integrable framework are detailed in Section \ref{sec:bethe}. Section \ref{sec:ope_coefficients} focuses on the general discussion of the problem of constraining linear combinations of squared OPE coefficients and how to adapt it in \texttt{SDPB}. Our primary results, particularly concerning the bounds of the four-point function, are presented in Section \ref{sec:4pt}. We reflect on our results and future directions in Section \ref{sec:discussion}. Supplementary material, including superconformal block expansions, details on integrated correlators, and a selection of our results for the bounds, is provided in Appendices \ref{app:app1}, \ref{app:app2}, and \ref{app:app3} respectively. We provide additional data to a \texttt{Mathematica} notebook attached to this paper.

\section{Setup}\la{sec:setup}

Let us start with a concise introduction to the 1D CFT and the observables we will study in the paper. Correlation functions of the defect CFT are constructed as operator insertions on a 1/2 BPS Maldacena-Wilson loop~\cite{Drukker:2006xg} (for a recent review on Wilson lines as defects see \cite{Aharony:2023amq}). Thus, operators are separated by the following segment of Wilson-Maldacena operator,
\beq\label{eq:segment}
{\cal W}_{t_1,t_2} \equiv \operatorname{P}\exp\left(\int_{t_1}^{t_2} dt\,( i A_{\mu}(t) \dot x_\mu
+ \Phi_6(t)|\dot x|)\right)\;,
\eeq
where $x^{\mu}(t)$ parametrises a line segment in 4D space, e.g. $x^{\mu} = (t,0,0,0)$, and the covariant derivative is $D_\mu=\d_\mu-i A_\mu$. 
In these conventions, all fields are $N\times N$ Hermitian matrices. 
In \eqref{eq:segment}, $\Phi_6$ is one of the six scalar fields of the $\mathcal{N}$=4 SYM theory defining a fixed polarization, and  $\operatorname{P}$ is the ``later-first'' path-ordering $\operatorname{P}(O_1(t_1)O_2(t_2))\equiv \Theta_{t_2>t_1}O_2(t_2)O_1(t_1)+
\Theta_{t_2<t_1}O_1(t_1)O_2(t_2)$.
The operator insertions are then constructed as~\cite{Drukker:2006xg}
\begin{equation}\label{eq:npoint}
    \left\langle\left\langle O_{1}\left(t_{1}\right) O_{2}\left(t_{2}\right) \cdots O_{n}\left(t_{n}\right)\right\rangle\right\rangle
    \!\equiv 
\frac{\langle
    \operatorname{Tr}
   \operatorname{P}{ O}_1(t_1)
 \,{ O}_{2}(t_{2})
    \ldots 
 O_n(t_n)\,\mathcal{W}_{-\infty,\infty}
    \rangle}{\langle \mathcal{W}_{-\infty, \infty}\rangle} ,
\end{equation}
where $O_i$ are renormalised composite fields inserted along the line, and transforming in the adjoint representation of the gauge group.

These correlation functions admit a 1D conformal symmetry,
\beq
 \left\langle\left\langle O_{1}\left(t_{1}\right) O_{2}\left(t_{2}\right) \cdots O_{n}\left(t_{n}\right)\right\rangle\right\rangle = \prod_{i=1}^n \left| \frac{\partial t'_i }{\partial t_i} \right|^{\Delta_i} \;  \left\langle\left\langle O_{1}\left(t_{1}'\right) O_{2}\left(t_{2}'\right) \cdots O_{n}\left(t_{n}'\right)\right\rangle\right\rangle ,
\eeq
where $t\rightarrow t'  = \frac{a t + b }{c t + d}$, $ ad - b c = 1$ is any 1D conformal transformation  which preserves the cyclic order of the points on the line. The possibility to consider cyclically related orders comes from the embedding in the 4D theory: using a 4D conformal transformations we can relate these correlators to insertions on a circular Maldacena-Wilson loop (divided by the vev of the empty circle), which makes the cyclic invariance manifest. Using standard arguments one can then constrain the form of 2 and 3-point functions, i.e. choosing a normalisation,
\beqa
\big\langle\big\langle O_{i}\left(t_{1}\right) \bar O_{j}\left(t_{2}\right) \big\rangle\big\rangle &=&\frac{\delta_{ij}}{ |t_{12}|^{2\Delta_i}}\;, \\
\big\langle\big\langle O_{i}\left(t_{1}\right) O_{j}\left(t_{2}\right) O_{k}\left(t_{3}\right) \big\rangle\big\rangle &=&\frac{C_{ijk}}{|t_{12}|^{\Delta_i+\Delta_j-\Delta_k} |t_{13}|^{\Delta_i+\Delta_k-\Delta_j} |t_{23}|^{\Delta_j+\Delta_k-\Delta_i}} , \;\;\; t_3 > t_2 > t_1 ,\nonumber
\eeqa
where, importantly, the OPE coefficients $C_{ijk}$ depend on the order of insertions of the operators on the line, since the allowed conformal transformations are only the ones that preserve such order.

The 1D CFT so defined admits a OSp($4^{\ast} $|$ 4$) superconformal symmetry~\cite{Liendo:2016ymz}. The most basic protected fields are the five scalars $\Phi_M$, $M=1,\dots, 5$ with polarisation orthogonal to the one running on the line. They have fixed dimension $\Delta=1$, and are the top component of a short superconformal multiplet of protected operators named $\mathcal{B}_1$ \cite{Liendo:2018ukf}. The theory also contains protected multiplets $\mathcal{B}_n$, with top component of dimension $\Delta = n$, and an infinite number of non-protected multiplets denoted as  $\mathcal{L}^{\Delta}_{s , [a,b] }$, with subscripts labelling the charges of the top component under global symmetries $SO(3) \times Sp(4)$, which have a nontrivial scaling dimension depending on the coupling.

The simplest bootstrap setup, studied in many works  e.g. \cite{Giombi:2017cqn, Liendo:2018ukf, Ferrero:2021bsb, Cavaglia:2021bnz} considers the 4-point function of the simplest protected scalars as follows\footnote{For definiteness, we choose identical polarisations for the four fields. Different R-symmetry channels can be explicitly related to this one using supersymmetry, see \cite{Liendo:2018ukf}. }
\beq\label{eq:4pt0}
\langle \langle \Phi_M(t_1) \Phi_M(t_2) \Phi_M(t_3) \Phi_M(t_4) \rangle \rangle = \frac{G(x) }{t_{12}^2 t_{34}^2} , \;\;\; M \in\left\{ 1,\dots,5\right\}, 
\eeq
where the 1D cross ratio is given by
\beq
x = \frac{t_{12} t_{34} }{t_{13} t_{24}}, \;\
\;\; t_{ij} \equiv t_i - t_j\;.
\eeq
Thanks to the possibility to relabel the points cyclically $(1234)\rightarrow(2341)$, this 4-point function of identical operators satisfies the crossing equation
\beq\label{crossG}
(1-x)^2 G(x) = x^2 G(1-x)\;.
\eeq
The amplitude $G(x)$ will be a major focus of this work. It was shown in \cite{Liendo:2018ukf} that it can be rewritten in terms of a simpler reduced amplitude $f(x)$ as follows
\beq\begin{split}\la{pt4}
{G}(x) = \mathbb{F} \;x^2 +  (2 x^{-1} - 1)f(x) -\left(x^2 - x +1\right)f'(x)\; ,
\end{split}\eeq
where $\mathbb{F}$ is an explicit function of the 't Hooft coupling $g$ (see appendix \ref{app:app2}). The reduced correlator also satisfies crossing in the following form
\beq\label{crossf}
(1-x)^2  f(x) + x^2 f(1-x) = 0\;,
\eeq
and it can be written in terms of the OPE decomposition \eqref{eq:OPEfusion} as
\begin{equation}\label{eq:reducedOPE}
f(x) = F_{\mathbb{I}}(x) + { {C^2_{\rm BPS} \,  {F}_{\mathcal{B}_2}(x)}}  + \sum_{n } { {C^2_{n} \,  {F}_{{\Delta_n}}(x)}} \; ,
\end{equation}
with superconformal blocks $F_{\bullet}$ given by \eqref{superblocks}. In \eqref{eq:reducedOPE}, $\mathbb{I}$ is the identity an $C^2_{\rm BPS}$ is the structure constant corresponding to the $\mathcal{B}_2$ block. The latter is a known function of the coupling computed both using supersymmetric localisation \cite{Liendo:2018ukf} as well as later reproduced using only integrability 
 arguments \cite{Cavaglia:2022qpg}. Its value is reported in appendix \ref{app:app2}. 
 
\begin{figure}[h]
    \centering
    \includegraphics[width=.8\columnwidth]{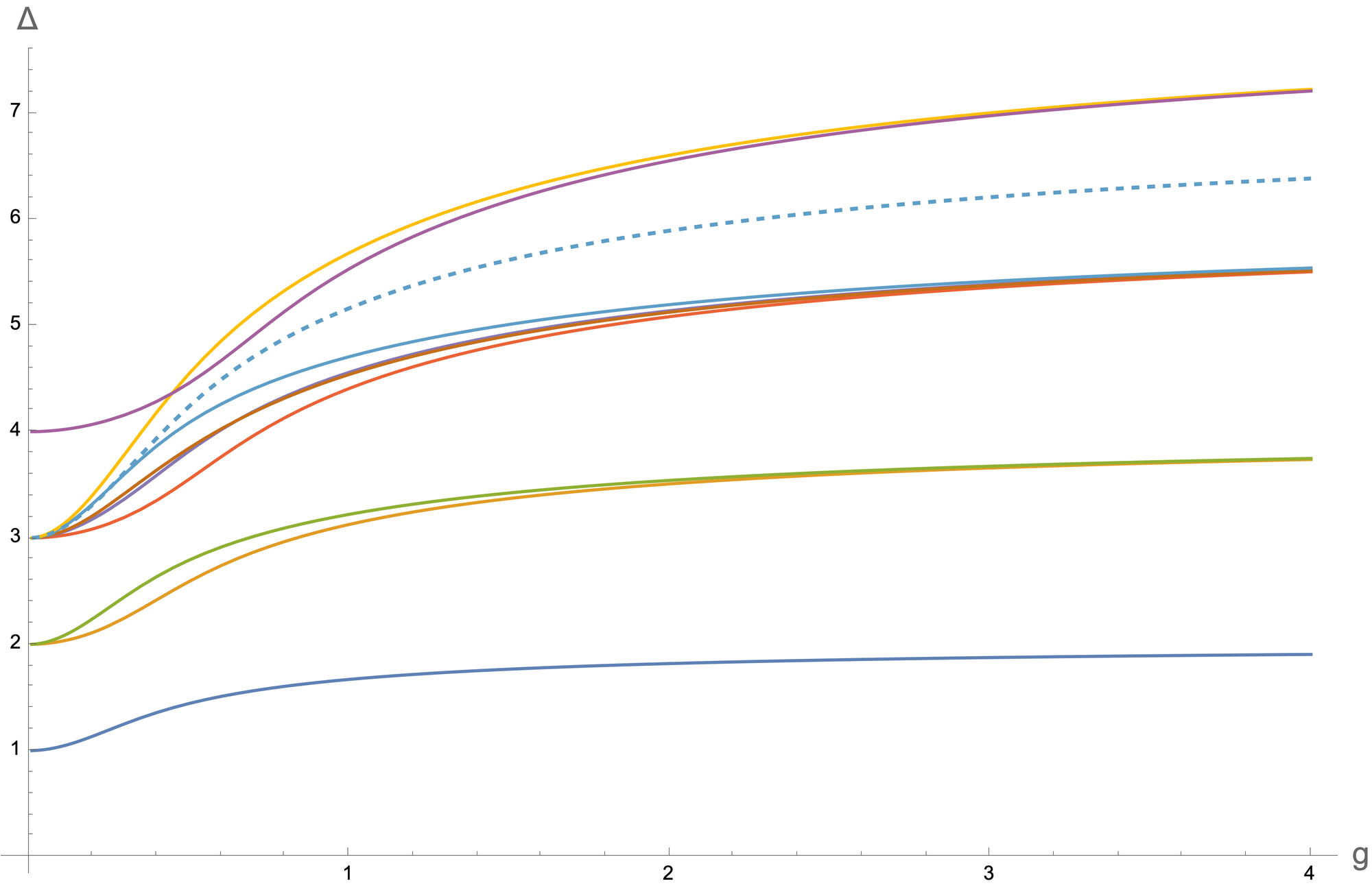}
    \caption{The first 10 spectral levels as functions of the 't Hooft coupling. 
    The dashed spectral line corresponds to the dimension of the state $O_7$ that does not contribute to four-point function \eqref{eq:4pt0}. See section \ref{sec:parity_symmetry} for details.}
    \label{fig:spectrum10}
\end{figure}

 The infinite sum in the OPE is over the full spectrum of the $\mathcal{L}^{\Delta}_{0,[0,0]}$ non-protected multiplets, with OPE coefficients 
\beq\label{cn}
C_n \equiv C_{\Phi_{M} ,\; \Phi_{M} ,\;\mathcal{L}_{0,[0,0]}^{\Delta_n}}\,.
\eeq
Using crossing symmetry, we get the basic bootstrap constraint
\beq\la{eq:crossing0}
\mathcal{G}_{\mathbb{I}}(x)+C_{\rm BPS}^2\mathcal{G}_{\mathcal{B}_2}(x) +\sum_n C^2_n \mathcal{G}_{\Delta_n}(x) = 0\;,
\eeq
with $\mathcal{G}_{\bullet}(x) \equiv (1-x)^2 F_{\bullet}(x) + x^2 F_{\bullet}(1-x)$.
The non-protected spectrum is accessible with the Quantum Spectral Curve method, developed for this problem in \cite{Grabner:2020nis, Julius:2021uka,Cavaglia:2021bnz,NikaJuliusFuture}. The dimensions of the first 10 states, which will be used in this paper as an input for Bootstrability were computed in \cite{Cavaglia:2021bnz} and shown in figure \ref{fig:spectrum10}.

The correlation function \eqref{eq:npoint} also satisfies two additional integrated constraints, which can be written as\footnote{While written in a different form, these are the same identities as in \cite{Cavaglia:2022qpg}. The subtraction of the two leading terms of $f(x)$ at $x \sim 0$ is convenient for writing a  compact convergent integral in both cases.}
\begin{align}
\int_{0}^{1/2}
     \left(f(x) - x +\frac{C^2_{\rm BPS}(g)}{2} x^2\right)\mu_a(x) dx +\mathcal{K}_a(g) = 0\;\;,\;\;a=1,2\; ,\label{eq:constr12}
\end{align}
where $\mu_a(x)$ are some simple rational functions of $x$ and $\mathcal{K}_a$ are exact functions of the coupling defined in the Appendix \ref{app:app2}.
Such constraints were introduced in \cite{Cavaglia:2022qpg} which we refer to for details. There, it was shown that incorporating these constraints in the numerical bootstrap setup leads to huge gains in precision. Later, in \cite{Drukker:2022pxk,Cavaglia:2022yvv}, a complete proof was provided exploiting conformal perturbation theory techniques.

The Bootstrability program developed until now was used to put bounds on individual OPE coefficients. In the rest of this paper, first we present new consequences of the existence of a discrete symmetry in the 1D CFT, which further improve our bounds. We then go on to apply the bootstrap method to get bounds directly on the full 4-point function at generic value of the cross ratio $x \in [0,1]$.

\section{Parity Symmetry}\la{sec:parity_symmetry}
In this section we introduce the QFT description of a particular $\mathbb{Z}_2$-parity symmetry. Whereas the discrete symmetries as such in the context of defect CFTs are not new (see e.g.~\cite{Billo:2013jda,Homrich:2019cbt,Bianchi:2020hsz}), its concrete realisation in the ${\cal N}=4$ SYM and the way to decode the parity charge from the integrability data is the main focus of this and the following sections.

In order to motivate the parity symmetry let us start from examples of some states, which can be obtained from the one-loop Hamiltonian of \cite{Correa:2018fgz}. Let us consider the lowest-lying seven states built only in terms of scalar fields at weak coupling. Labelling them as $O_n$ corresponding to the magnitude of their scaling dimensions at weak coupling (at strong coupling the order is different as there are some level crossings), they have the following form at the leading order in perturbation theory~\cite{Correa:2018fgz,NikaJuliusFuture}:\footnote{We thank Julius Julius and Nika Sokolova for communications.}
\beqa
O_1 &=& \Phi_6 , \\
O_{2,3} &=&  \sqrt{5}\Phi_6^2 \pm \Phi^{\perp}_i\Phi_i^{\perp} \;, \\
O_{4,5,9} &=& \Phi_6^3 + \alpha_a \Phi^{\perp}_i \Phi_6 \Phi_i^{\perp}+ \beta_a (\Phi_6\Phi_i^{\perp}  \Phi_i^{\perp}+  \Phi_i^{\perp}  \Phi_i^{\perp}\Phi_6)\;,\\
O_7&=&\Phi_6\Phi^{\perp}  \Phi^{\perp}-  \Phi^{\perp}  \Phi^{\perp}\Phi_6\; ,
\eeqa
where $\alpha_a$, $\beta_b$ are certain coefficients that can be found by diagonalising the one-loop Hamiltonian of \cite{Correa:2018fgz}. We omitted from this list the states $O_6, O_8$ since they are not built out of only scalar fields at tree level.
We see that, unlike the case of the closed spin chain describing local operators, there is no cyclic permutation symmetry here. At the same time, the one-loop states above have a definite symmetry under reverting the order of the constituent fields. From the explict form of the operators it is clear that one can associate a parity  charge ${\mathbb P}=\pm 1$  to each eigenstate of this transformation. The states $O_n$ above have $\mathbb{P} = 1$ for $n\neq 7$, and $\mathbb{P} = -1$ for the state $O_7$.  The question is if this symmetry persists beyond one loop. We answer this affirmatively in the next section.

\subsection{Charge conjugation and Parity of the operators non-perturbatively}\la{sec:pariti}
The Lagrangian of ${\cal N}=4$ SYM is invariant under the charge-conjugation transformation, whereas all fundamental fields, which are $N\times N$ matrices with $N$ the rank of the gauge group, are transformed as\footnote{We are grateful to Shota Komatsu for discussing this point.}
\beq\label{eq:chargeconj}
F \to -F^T .
\eeq
This transformation preserves the $SU(N)$ commutation relations and in general maps one $SU(N)$ representation to its conjugate (e.g. fundamental to anti-fundamental). All the fields in ${\cal N}=4$ SYM are in the adjoint representation, which is self-conjugate, i.e. the above transformation (\ref{eq:chargeconj}) is an automorphism. Therefore,  this substitution leaves the Lagrangian invariant. Let us see which effect this symmetry has on the observables.

For the correlator of any number of ordinary (non supersymmetric) Wilson lines, we get an invariance under orientation change of all of them simultaneously
\beq
\Big\langle \text{Tr}\text{Pexp}\left(i\int_{\gamma} A_{\mu} dx^{\mu}\right)   \Big\rangle = 
\Big\langle \text{Tr}\text{Pexp}\left(i\int_{\gamma} (- A_{\mu} )^T dx^{\mu}\right)   \Big\rangle  =
\Big\langle\text{Tr}\text{Pexp}\left(i\int_{-\gamma} A_{\mu} dx^{\mu} \right)   \Big\rangle\;.
\eeq
Above, the first equality comes from the invariance of the SYM action under the charge conjugation symmetry \eq{eq:chargeconj}. 
The second equality then rewrites this using the property of the trace and  of the path-ordered exponential, and the net effect is changing the orientation of the loop. 

In the case of the Wilson-Maldacena loop, one has an additional
$\Phi_6 |dx|$ term in the exponent, which now  changes sign under the symmetry:
\beqa
&&\Big\langle \text{Tr}\text{Pexp}\left(\int_{\gamma} (i A_{\mu} dx^{\mu} + \Phi_6 |dx| )\right)  \Big\rangle = 
\Big\langle \text{Tr}\text{Pexp}\left(\int_{\gamma}( i(- A_{\mu})^T dx^{\mu} + (-\Phi_6^T) |dx| )\right)   \Big\rangle =\nonumber \\
=
&&\Big\langle\text{Tr}\text{Pexp}\left(\int_{-\gamma} ( i A_{\mu} dx^{\mu} - \Phi_6 |dx|)\right)  \Big\rangle\;.
\eeqa
Therefore, under the charge conjugation symmetry we get, in addition to the orientation reversal, the direction in R-space changed. This of course can be compensated by the fact that the action is also invariant under the discrete $\Phi_6\to -\Phi_6$ symmetry, which allows to change the sign back.

Now consider the case with insertions of operators along the contour at $x(t_1)$, $x(t_2)$ etc. In this case, after performing the combined symmetries of charge conjugation \emph{and} the sign-flip of $\Phi_6$, we get
\beqa
&&\Big\langle \text{Tr}\left[\dots    O_2(t_2){\cal W}_{t_1,t_2}O_1(t_1) \dots\right] \Big\rangle = \Big\langle \text{Tr}\left[\dots   O_1^c(t_1){\cal W}_{t_2,t_1}O_2^c(t_2)   \dots \right]\Big\rangle,
\eeqa
where the Wilson line in the r.h.s. is running in the opposite direction. 
As the path has changed orientation, the operators have reversed their order along the contour, and moreover the operators are replaced by their ``charge-conjugated'' version. Representing an operators as a products of adjoint fields with any number of covariant derivatives $f_i$, their charge conjugate is given by
\beq\la{Oc}
O = ( f_1 f_2 \dots f_n) \quad\rightarrow\quad O^c = \left.(-1)^n f_n \dots f_2 f_1\right|_{\Phi_6 \to - \Phi_6
}\;.
\eeq

Notice that the change in the orientation of the contour can be undone easily in the case of two insertions along a straight line, as we can use the reflection symmetry $x_4 \rightarrow -x_4$, for the coordinate $x_4$ along the contour. Thus, the charge conjugation \eq{Oc} in the combination with the reflection $x_4 \rightarrow -x_4$ should be a $\mathbb{Z}_2$ symmetry commuting with the dilatation operator which we denote by $O\to O^P$. From that we see that all operator insertions can be chosen to be either odd or even under the transformation. This allows to organize the primary operators of the 1D CFT in two sectors corresponding to their charge ${\mathbb P}=\pm 1$ defined by $O^P={\mathbb P}\; O$. From now on, we will use a basis of operators of definite parity. 

Multi-point correlation functions are in general not invariant under this parity symmetry but transform in a simple way 
\beq\label{eq:parityprop}
\langle\langle O_1(t_1) \dots O_n(t_n) \rangle\rangle = {\mathbb P}_1\dots{\mathbb P}_n \langle\langle O_n(-t_n) \dots O_n(-t_1) \rangle\rangle\; ,
\eeq
where $t$ is the component of the 4D coordinate parametrising the line.

\subsubsection{Implications for OPE coefficients}\label{sec:OPEimpli}
Let us see the consequences of (\ref{eq:parityprop}) for a 3-point function.\footnote{see also appendix K of \cite{Homrich:2019cbt}  for a systematic discussion of parity in the context of CFT$_1$'s.  
} In this case, the dependence on the coordinates is controlled by the conformal symmetry, but the order of operators along the line does matter, with only cyclically related orderings giving the same OPE coefficient (it is convenient to think of the points as lying on a circle, which is obtained from the line after a conformal transformation). Applying the general formula \eq{eq:parityprop}, we get
\beq\la{parityC}
C_{O_1O_2O_3} = C_{O_3O_2O_1} \; {\mathbb P}_1 {\mathbb P}_2 {\mathbb P}_3\; .
\eeq
This relation reduces the number of unknown parameters in the conformal bootstrap procedure, as we can always fix a particular ordering and consider the corresponding OPE coefficient as a parameter, characterising the other possible cyclic ordering.

In the special case where two operators are equal, we get by cyclicity 
\beq
C_{O_1O_1O_2} = C_{O_2O_1O_1}\;,
\eeq
but also at the same time from \eq{parityC} we have
\beq\la{zeroC}
C_{O_1O_1O_2} = C_{O_2O_1O_1} {\mathbb P}_2 {\mathbb P}_1^2 = C_{O_2O_1O_1} {\mathbb P}_2= C_{O_1O_1O_2} {\mathbb P}_2\;,
\eeq
meaning that \textit{only} $\mathbb{P}$-even operators can appear in the OPE of two equal operators, which is the case relevant for the correlator (\ref{eq:4pt0}) studied in this paper. This allows to remove some part of the spectrum, improving the bounds for the remaining OPE coefficients. In particular, one can remove the scaling dimension of $\mathbb{P}$-odd operators since they have vanishing 3-point function\footnote{Notice that 
 here we are making statements about scalar operators, or alternatively about individual components of R-symmetry representations. We need to be careful when different R-symmetry indices are combined. For instance, in the OPE of $\mathcal{B}_2 \times \mathcal{B}_2$ flow the $\mathbb{P}$-\emph{even} operators  of $\mathcal{L}^{\Delta}_{[0,0]}$ and $\mathcal{L}^{\Delta}_{[0,2]}$ multiplets, and the 
 $\mathbb{P}$-\emph{odd} operators  of the $\mathcal{L}^{\Delta}_{[2,0]}$ one. This boils down to the fact that $[2,0]$ is the antisymmetric representation, which results in an additional sign. In short, from \eq{parityC} we find $C_{\mathcal{B}_2,\mathcal{B}_2,\mathcal{L}^{\Delta}_{[2,0]}}= - {\mathbb P}_{\mathcal{L}^{\Delta}_{[2,0]}}C_{\mathcal{B}_2,\mathcal{B}_2,\mathcal{L}^{\Delta}_{[2,0]}}$and $C_{\mathcal{B}_2,\mathcal{B}_2,\mathcal{L}^{\Delta}_{[0,2]}}= + {\mathbb P}_{\mathcal{L}^{\Delta}_{[0,2]}}C_{\mathcal{B}_2,\mathcal{B}_2,\mathcal{L}^{\Delta}_{[0,2]}}$.
 }
\begin{equation}\label{Codd0}
C_{O_1O_1O_{\mathbb{P}-odd}}=0\;.
\end{equation}

The vanishing of some OPE coefficients at several orders at strong coupling was noticed in the context of the analytical bootstrap~\cite{Ferrero:2021bsb}, and it was conjectured that this fact might be valid at all orders, at least in strong coupling perturbation theory. The exact parity symmetry of the theory we have discussed gives a simple explanation of the phenomenon, and shows that the vanishing of these coefficients is valid non-perturbatively.  In section \ref{sec:bethe} we show how to determine the parity charge of states and thus determine precisely which OPE coefficients are vanishing, bypassing building these operators explicitly at weak coupling from the ABA/QSC data.
\subsubsection{Complex conjugation}\label{sec:complex}
Let us also revise the  argument about the properties of the OPE coefficients under complex conjugation. The action of ${\cal N}$=4 SYM is real, and the fields are Hermitian as $SU(N)$ matrices.

So we have, for the Wilson-Maldacena loop itself:
\beqa
\overline{\Big\langle \text{Tr}\;\text{Pexp}\left(\int_{\gamma} (i A_{\mu} dx^{\mu}+\Phi_6|dx|) \right)   \Big\rangle} &=& 
\Big\langle \text{Tr}\[\text{Pexp}\left(\int_{\gamma} (i A_{\mu}  dx^{\mu}+\Phi_6|dx|)\right)   \]^\dagger \Big\rangle\\
\nonumber &=& 
\Big\langle \text{Tr}\;\text{Pexp}\left(\int_{-\gamma} (i A_{\mu}  dx^{\mu}+\Phi_6|dx|)\right)    \Big\rangle
\; ,
\eeqa
which means that complex conjugation changes the contour orientation. In order to restore the initial orientation we again perform the same reflection as in section~\ref{sec:pariti}. Looking at $n$-point insertions along a WM line, we have 
\beq\label{eq:intermediate}
\overline{\langle\langle
O_1(t_1)\dots O_n(t_n)
\rangle\rangle}
=
\langle\langle
\bar O_n(-t_n)\dots \bar O_1(-t_1)
\rangle\rangle\;,
\eeq
where $\bar O$ is hermitian conjugation with the reflection along the time around the point where the operator sits. This is in agreement with the reflection-positivity principle, ensuring the positivity of 2-point functions $\langle\langle O\bar O\rangle\rangle > 0$ (see \cite{Kravchuk:2021kwe} for a recent review)\footnote{The $2$-point functions of Hermitian conjugate operators is not necessary positive e.g. for a derivative of a scalar.}.

Restoring the normal contour orientation using charge conjugation, from (\ref{eq:intermediate}) we get
\beq
\overline{\langle\langle
O_1(t_1)\dots O_n(t_n)
\rangle\rangle}
=
{\mathbb P}_1\dots{\mathbb P}_n
\langle\langle
\bar O_1(t_1)\dots \bar O_n(t_n)
\rangle\rangle\; .
\eeq
This implies, in the case of 3-point functions, 
\beq\label{eq:OPEcomplex1}
\bar C_{O_1 O_2 O_3} ={\mathbb P}_1{\mathbb P}_2{\mathbb P}_3 C_{\bar O_1 \bar O_2 \bar O_3 }\;. 
\eeq

\paragraph{Reality Implications on OPE decomposition. } 
Let us analyze the consequences of these reality properties on OPE coefficients. 

Assuming the standard normalisation of the 2-point function $\langle \langle \bar{O}_i 
 O_j \rangle \rangle = \delta_{ij} |t_{ij}|^{-2 \Delta_i}$ does not fix the basis of operators completely, as one can still multiply the operators by an arbitrary phase factor. 
 This can be used to impose reality of some structure constants as we discuss below.
Instead of the individual structure constants, however, let us consider the combinations that appear in the OPE decomposition of a 4-point function, i.e.,
\beq
\langle \langle {O}_1 O_2 O_3 O_4 \rangle \rangle =\sum_n C_{O_1O_2 O_n} C_{ \bar{O}_n O_3 O_4} \mathcal{F}_{1234, \Delta}(x) \times \left(\texttt{kinematical factors}\right),
\eeq
with $\mathcal{F}_{1234, \Delta}(x)$ denoting the conformal blocks, and where we omitted the precise kinematical factors since they are not consequential here. For simplicity, in the rest of this section, we will omit completely kinematics as well as conformal blocks, and write only schematic OPE expansions. 

We see that the relevant quantities are products  $C_{O_1O_2 O_n} C_{ \bar{O}_n O_3 O_4}$,
which are invariant under redefining the exchanged operators by phase factors. 
As a consequence of the parity and reality  discussed above for some choice of external operators
 these combinations will have specific reality properties. Indeed, from (\ref{eq:OPEcomplex1}), we immediately find
\beq
C_{O_n O_1 O_2}  
 C_{ \bar{O}_n{O}_3 {O}_4 } = \mathbb{P}_1 \mathbb{P}_2 \mathbb{P}_n 
 C_{O_n O_1 O_2}  
  \bar{C}_{ {O}_n \bar{O}_3 \bar{O}_4 }\;. 
\eeq
Let us consider some specific simple cases in turn. 
\begin{itemize}
\item When $O_3 = \bar O_1$, $O_4= \bar O_2$. This is the situation we encounter in the OPE decomposition of $$\langle \langle \underbrace{O_1 O_2 }_{\text{OPE}} {\bar O_1 \bar O_2 } \rangle \rangle ,$$ 
and the exchanged operators come with the coefficient with positivity determined by the parity charges:
\beq
\langle \langle \underbrace{O_1 O_2 }_{\text{OPE}} {\bar O_1 \bar O_2 } \rangle \rangle = \sum_n C_{O_n O_1 O_2}  
 C_{ \bar {O}_n\bar {O}_1 \bar {O}_2 }   =\sum_n \mathbb{P}_n \mathbb{P}_1 \mathbb{P}_2 | C_{O_n O_1 O_2}  |^2  .
\eeq
\item Another interesting case is a correlator with $O_3 = \bar O_2$, $O_4= \bar O_1$. In this case the OPE decomposition in the $s$-channel leads to exchanged operators with coefficient which is \emph{always} strictly positive:\beq\langle \langle \underbrace{O_1 O_2 }_{\text{OPE}} {\bar O_2 \bar O_1 } \rangle \rangle = \sum_n 
C_{O_n O_1 O_2}  
 C_{ \bar {O}_n \bar {O}_2 \bar {O}_1 }   = \sum_n | C_{O_n O_1 O_2}  |^2  .
\eeq
\item An interesting simpler case is when $O_2 = O_1$, $O_3 = O_4 = \bar O_1$. In this case the $s$-channel OPE decomposition gives
\beq\langle \langle \underbrace{O_1 O_1 }_{\text{OPE}} {\bar O_1 \bar O_1 } \rangle \rangle =\sum_n 
C_{O_n O_1 O_1}  
 C_{ \bar {O}_n \bar {O}_1 \bar{O}_1 }   = \sum_{n: \;\; \mathbb{P}_n = 1} | C_{O_n O_1 O_1}  |^2   ,
\eeq
where moreover we have used the result that $C_{O_n O_1 O_1} = 0$ when $\mathbb{P}_n$ has odd parity, deduced in section \ref{sec:OPEimpli}. 
\item The decomposition of the same correlator in the $t$-channel gives
\beq\langle \langle O_1 \underbrace{O_1 
 \bar O_1 }_{\text{OPE}} \bar O_1  \rangle \rangle =\sum_n 
C_{O_n O_1 \bar O_1}  
 C_{ \bar {O}_n\bar {O}_1 {O}_1 }   = \sum_{n} \mathbb{P}_n | C_{O_n O_1 \bar O_1 }  |^2   .
\eeq

\end{itemize}
Such reality and positivity properties allow to study  certain systems of correlators with the methods of the unitary conformal bootstrap. A careful inspection of the reality properties allows to reduce the number of unknown real parameters in the optimization procedures used in the bootstrap approach, such as the ones based on \texttt{SDBP}. 

For example, a system of four correlation functions was studied in \cite{Liendo:2018ukf}. In an upcoming work with Julius Julius and Nika Sokolova~\cite{MultiC}, we reexamine this problem from the point of view of Bootstrability and of the parity symmetry discussed here. We will see that the parity symmetry leads to a significant reduction of the search space for OPE coefficients, roughly by half, leading to narrower bounds.

\subsection{Parity at Strong Coupling}
In the next section, we show how one can exploit the integrability description of a state (in terms of Bethe equations arising at weak coupling) to determine the value of its parity charge $\mathbb{P}$. 

By computing the parity of several states with this method, and then tracking their evolution  to the strong coupling region, we noticed that in all cases we have studied\footnote{We thank Julius  Julius and Nika Sokolova for helping us in testing the integrability description of the parity charge and in finding this pattern.}
\beq\la{Pinf}
{\mathbb P}=(-1)^{\Delta_{\lambda=\infty}+R_1+R_2}\; ,
\eeq
for generic states in  $\mathcal{L}^\Delta_{0,[R_1,R_2]}$
multiplets.\footnote{This pattern likely has a simple interpretation considering the GFF description of the theory at infinite coupling, coming from the dual AdS$_2$ QFT of \cite{Giombi:2017cqn}. We thank Ant\'onio Antunes and Gabriel Bliard for comments.} The \eq{Pinf} was verified on $28$ states $\mathcal{L}^\Delta_{0,[0,0]}$,
$9$ states $\mathcal{L}^\Delta_{0,[0,1]}$,
$4$ states $\mathcal{L}^\Delta_{0,[0,2]}$ and
$3$ $\mathcal{L}^\Delta_{0,[2,0]}$ states i.e. for all states we have available at the moment. An analytic proof of \eq{Pinf} has a number of challenges from the QSC side: we only know the expression for ${\mathbb P}$ in terms of the weak coupling Bethe roots (see section~\ref{sec:bethe}) and furthermore analytic solution of QSC at strong coupling is still an open problem.
We postpone a more detailed scan of the states to the future~\cite{NikaJuliusFuture,MultiC}.

For the case of the single correlator described in this paper, we can make an interesting observation that all states with odd dimension at strong coupling decouple from the correlator due to  \eq{Codd0}. This is consistent with \cite{Ferrero:2021bsb,Ferrerotoaappear}, where the states with odd dimensions have not appeared in the analytic expressions for the $4$-point functions. At strong coupling there could be an independent argument leading to this result from the holographic description~\cite{AntunesUnpublished}.

The first of odd states is the state with $\Delta_{\lambda=\infty}=7$ (dashed line in figure \ref{fig:spectrum10}), which we previously included into the bootstrap procedure. In the next section we reanalyse our data excluding this state and discuss the implications for the precision of the result.

\begin{figure}[h]
    \centering
    \includegraphics[width=.8\columnwidth]{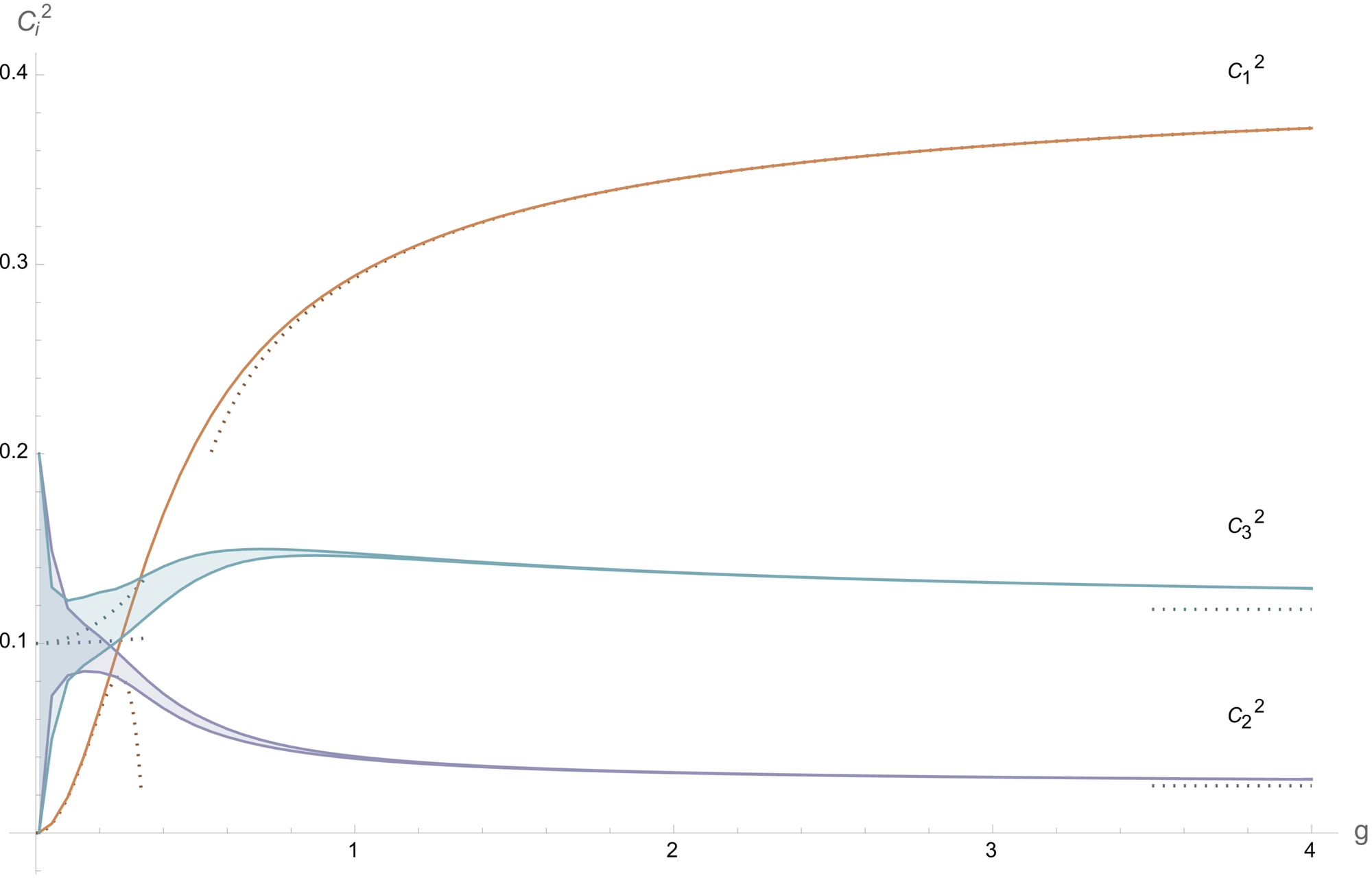}
    \caption{Updated results for the bounds of the first $3$ non-protected states structure constants from ${\cal L}_{0,[0,0]}^\Delta$ with two ${\cal B}_1$ BPS-states. The result appears similar to that from~\cite{Cavaglia:2022qpg} as the main precision improvement takes place in the area where the error was already small. See Fig.~\ref{figerror} for the error analysis.}
    \label{fig:C1}
\end{figure}

\begin{figure}[ht]
        \centering
        \subfloat{
            \includegraphics[width=.49\columnwidth]{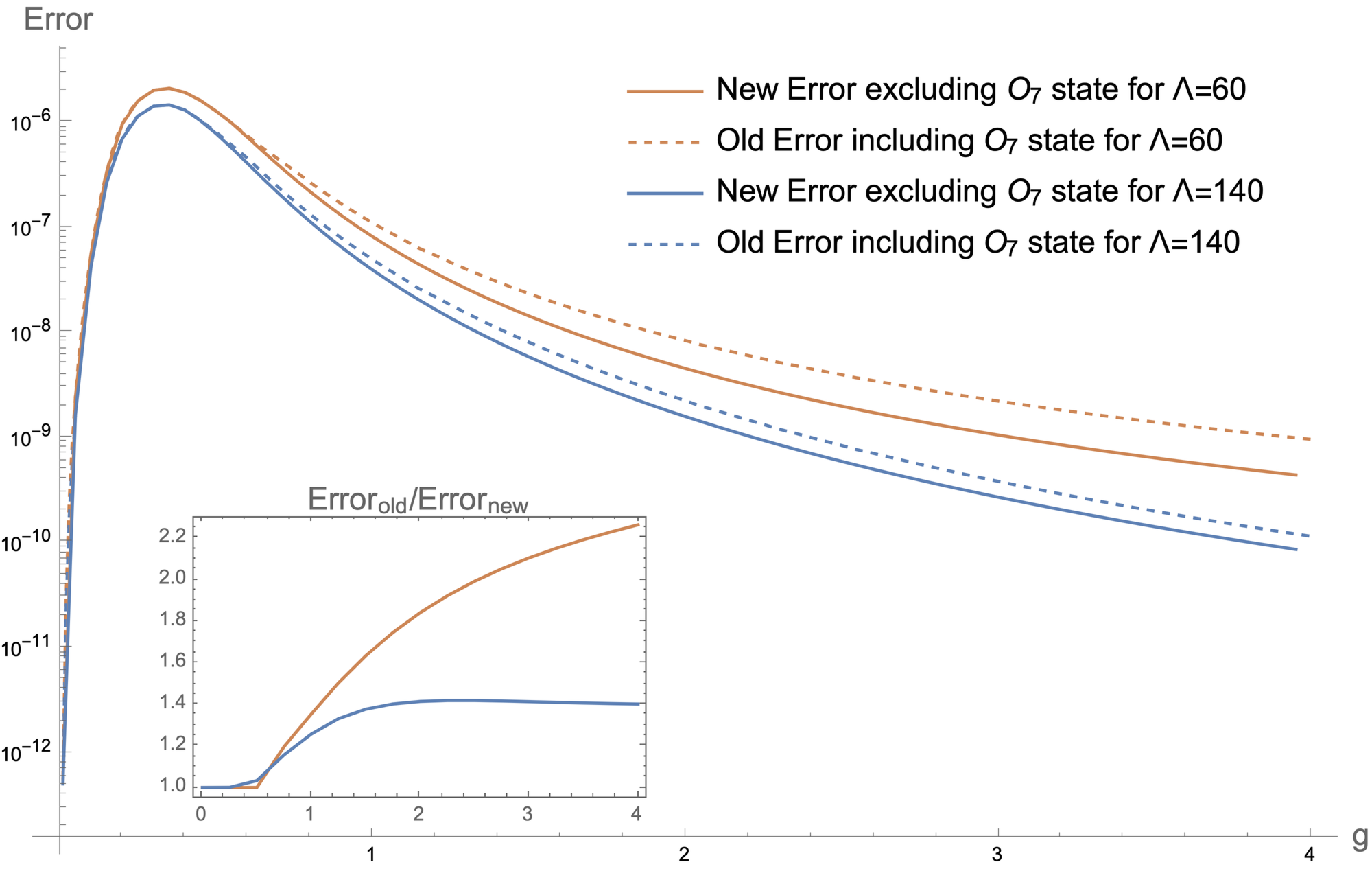}}
        \subfloat{
            \includegraphics[width=.49\columnwidth]{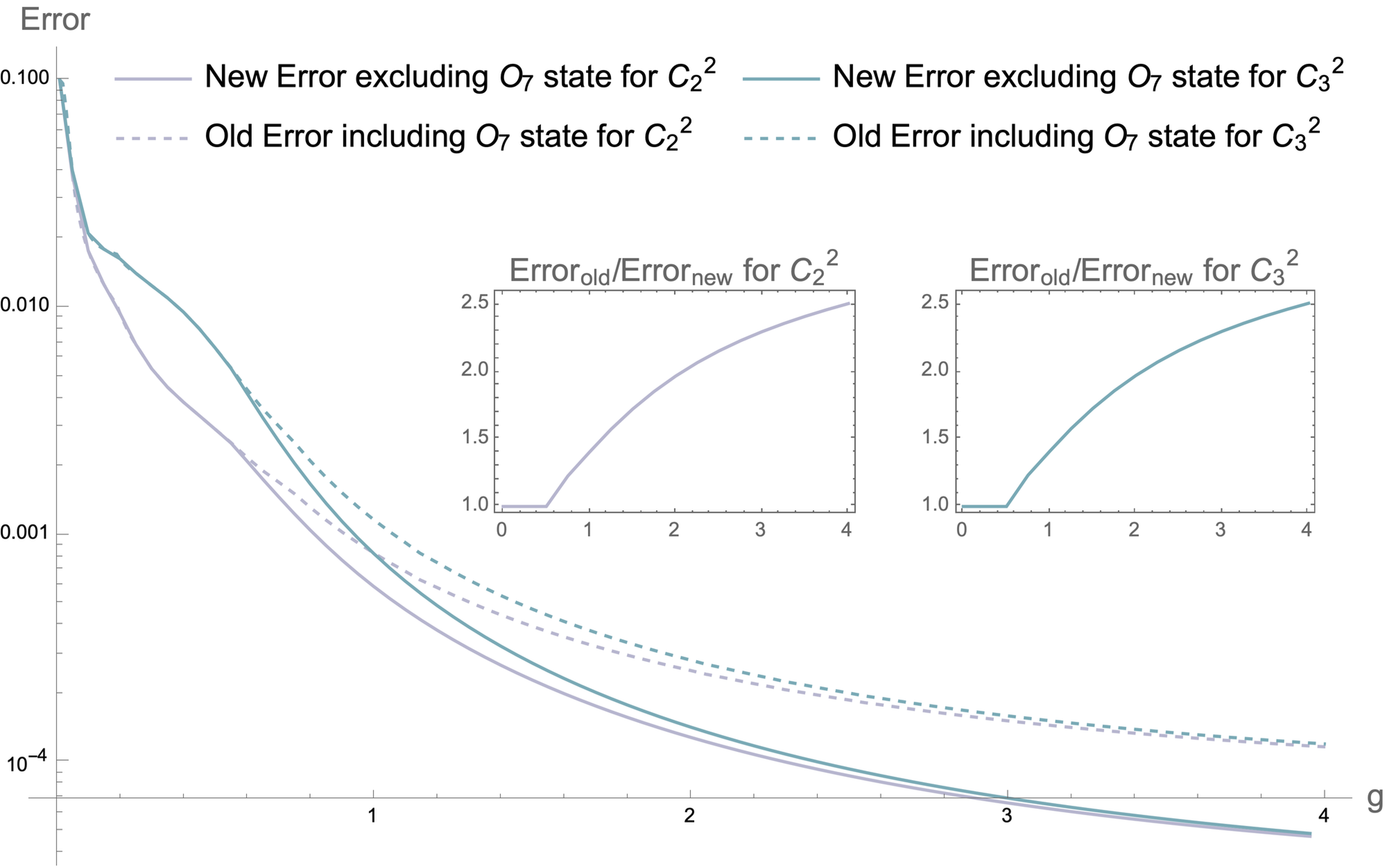}}
        \caption{Error for the structure constants in log-scale estimated as the half of the width of the bound obtained from the crossing equations. The error for  $C_1^2$ is well below $10^{-6}$ for most of the values of $g$ and the gain from excluding the odd state is up to $\sim 2.2$ for $\Lambda=60$ and $\sim 1.4$ for $\Lambda=140$ (Left). Improvement for excited states $C^2_2$ and $C^2_3$ for $\Lambda=60$ is by up to a factor $\sim 2.5$ (Right).}
        \label{figerror}
\end{figure}

\subsection{Gain from excluding odd states from conformal bootstrap}
As discussed below (\ref{zeroC}), states with $\mathbb{P} = -1$ do not contribute to the OPE expansion of two identical operators, as their OPE coefficients must vanish as in \eqref{Codd0}. As a result, we re-computed our numerical bounds from~\cite{Cavaglia:2022qpg}, including two integrated correlators and the spectrum of the first $10$ states, this time dropping out odd states from the analysis -- in practice, this means dropping the state $O_7$ which is the only $\mathbb{P}$-odd state in this part of the spectrum. We exploit the same \texttt{SDPB} implementation of \cite{Cavaglia:2022qpg} (we refer to this work for technical details) using a number of derivatives $\Lambda=60, 140$ to approximate the optimal functional. 

As the state $O_7$ appears at $\Delta = 3+9g^2+\dots$ at weak coupling, which is rather high up in the portion of the spectrum we consider, the effect of excluding it is not so enormous especially at weak coupling where there are $4$ more even states with the same bare dimension. However, at larger $g$ one gains a factor of $\sim 2$ or more of the precision. We noticed the same gain for both the ground state and the first two excited states (see Fig.\ref{fig:C1} and \ref{figerror}).

The gain in precision at strong coupling is expected. Indeed, it is known that when the spectrum is more sparse at strong coupling, the functional can become negative exploring the gap between different states~\cite{Cavaglia:2022qpg}. The more freedom the functional has to explore gaps, the better the precision. In our present case, excluding the state $O_7$, we open a big gap at strong coupling between states approaching $\Delta_{\lambda=\infty}=6$ and $\Delta_{\lambda=\infty}=8$ as presented in figure \ref{fig:spectrum10}. We expect this to be true also for higher states since, as explained in the previous sections, all the states approaching an odd integer at strong coupling can be excluded from the bootstrap setup.

We present a collection of the data in Appendix~\ref{app:app3} and the complete set in the Wolfram \texttt{Mathematica} notebook attached to this manuscript.

\section{Parity Operator from Integrability}\label{sec:bethe}

In the previous section we defined the parity charge, which is easy to compute when the operator is known explicitly in terms of the field content, e.g. when the one-loop dilatation operator is known as in the sector studied in \cite{Correa:2018fgz}. However, in general we work with operators in terms of integrability data such as Q-functions entering into the QSC. So for practical applications, especially in the sectors where the dilatation operator was not worked out explicitly even at one-loop, it is useful to have a way to determine the parity as a closed expression in terms of the integrability data.

\subsection{Warm-up example}
In order to determine the expression for the parity of a state in terms of the integrability data, we will use the description in terms of the Asymptotic Bethe Ansatz. While this is only an approximate picture, it becomes exact at weak coupling, where it can be connected with the exact QSC quantities. At the same time at the ABA level we can work with physically intuitive objects such as magnons, the scattering S-matrix and boundary scattering phase $R$. In this description the operators inserted along the Wilson line are represented as multi-magnon states on an open spin chain, with the line acting as boundary on which the magnons can reflect, see \cite{Correa:2012hh,Drukker:2012de,Correa:2018fgz}.

For simplicity, consider the simple case of scattering between $M$ scalar magnons (representing particular fields)  on an open spin chain of $L$ sites. A Bethe wave function is a certain special superposition of magnons with definite momenta $p_i$, which can be in all possible positions along the chain and scatter with each other and with the boundary in an integrable way. We can formulate the parity charge as the relative coefficient between terms in the wave-function with magnons in two positions related by a reflection across the middle of the chain -- e.g., concretely, the relative sign between the term with magnons on sites $i_1,\dots,i_M$ and the term with magnons at the positions $L-i_M,\dots,L-i_1$. The relative coefficient between these two configurations can be computed diagrammatically following figure~\ref{fig:magnons}, which depicts how the terms in the Bethe wave function are built using the scattering matrix $S(p_1, p_2)$ and the reflection matrix $R(p)$.
\begin{figure}
    \centering
    \includegraphics[scale=.5]{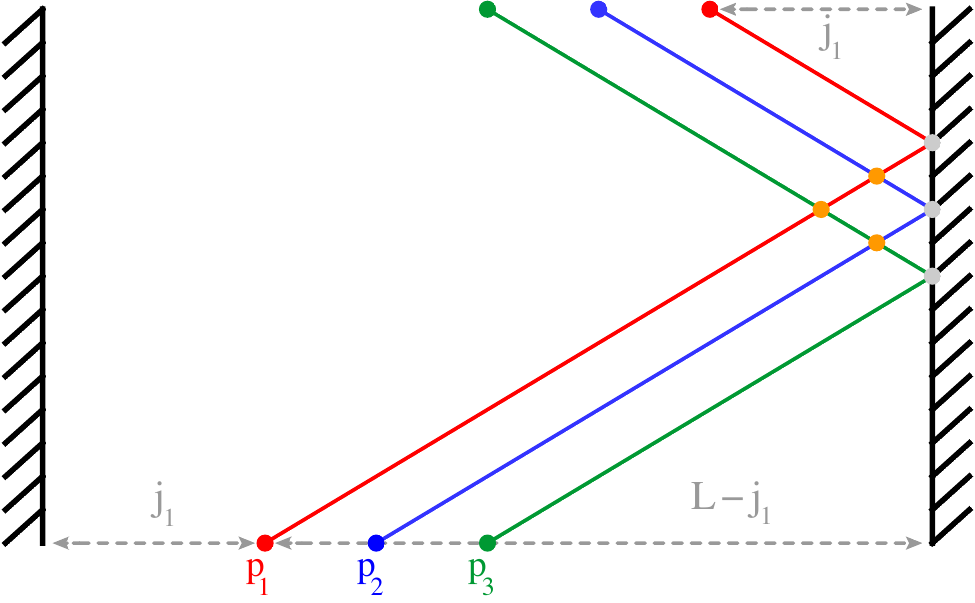}
    \caption{The relative coefficients between the two reflected configurations is determined by the particles free propagation factors by the distance $L$, scattering against all other particles (orange circles) as well as the boundary scattering phase (gray circles).}
    \label{fig:magnons}
\end{figure}
For the case of $3$ magnons, for this simple case, we get the following relative factor between the two configurations
\beq
{\mathbb P} = e^{i p_1 L}e^{i p_2 L}e^{i p_3 L} S(p_1,-p_2)S(p_1,-p_3)S(p_2,-p_3) R(p_1)R(p_2)R(p_3)\;,
\eeq
which generalizes easily to the $M$ magnon case as
\beq
{\mathbb P} = \prod_k  R(p_k) \,e^{i p_k L}\prod_{k<j} S(p_k,-p_j)\;.
\eeq
How can we be sure that ${\mathbb P}=\pm 1$? For that we have to write the quantization conditions on magnon momenta $p_k$ ensuring the periodicity of the wave-function, i.e. the Bethe equations
\beq\la{BAE}
1 = e^{2i p_k L}R^2(p_k)\prod_{j\neq k} S(p_k, p_j)S(p_k,-p_j)
\;\;,\;\;k=1,\dots, M\; ,
\eeq
then computing the product of all $M$ above equations, and using unitarity and parity $S(p,k)=1/S(k,p)$ and $S(p,-k)=S(k,-p)$ we arrive at
\beq\la{PPz}
{\mathbb P}^2 = \prod_{k=1}^M {\rm BAE}_k = 1\;,
\eeq
where ${\rm BAE}_k$ denotes the r.h.s. of \eq{BAE}.
We see that \eq{PPz} implies that ${\mathbb P}=\pm 1$. 

\subsection{Parity from QSC}
Note that due to the continuity, it is enough to determine the parity at weak coupling and then follow the solution continuously. The numerical solution of QSC is exactly doing that, by finding $\Delta$ based on the solution of QSC at slightly smaller $g$. 

At weak coupling the QSC simplifies considerably, so that many quantities become polynomials and the solution can be characterised in terms of the roots of these functions. Such roots are the Bethe roots at weak coupling, satisfying the rational-type Bethe equations arising at one loop from the QSC~\cite{Marboe:2014gma,NikaJuliusFuture}. Here let us assume that the Bethe roots are given. Such roots have to satisfy the Bethe ansatz equations, which are a consequence of the QQ-relations and read~\cite{Drukker:2012de,Correa:2012hh,NikaJuliusFuture}
\beqa
1&=&\prod_{j=1}^{M_2}\frac{u_{1,k}-u_{2,j}+i/2}{u_{1,k}-u_{2,j}-i/2}\equiv {\rm BAE}_{1,k}\\
\nonumber 1&=&\prod_{j\neq k}^{M_2}\frac{u_{2,k}-u_{2,j}+i}{u_{2,k}-u_{2,j}-i}
\prod_{j=1}^{M_1}\frac{u_{2,k}-u_{1,j}+i/2}{u_{2,k}-u_{1,j}-i/2}
\prod_{j=1}^{M_3}\frac{u_{2,k}-u_{3,j}+i/2}{u_{2,k}-u_{3,j}-i/2}\equiv {\rm BAE}_{2,k}\\
\nonumber 1&=&\prod_{j=1}^{M_2}\frac{u_{3,k}-u_{2,j}+i/2}{u_{3,k}-u_{2,j}-i/2}
\prod_{j=1}^{M_4}\frac{u_{3,k}-u_{4,j}-i/2}{u_{3,k}-u_{4,j}+i/2}
\equiv {\rm BAE}_{3,k}\\
\nonumber 1&=&\!\left(\frac{u_{4,k}-i/2}{u_{4,k}+i/2}\right)^{2L+1}\!
\prod_{j\neq k}^{M_4}\!\frac{u_{4,k}-u_{4,j}+i}{u_{4,k}-u_{4,j}-i}
\prod_{j=1}^{M_3}\!\frac{u_{4,k}-u_{3,j}-i/2}{u_{4,k}-u_{3,j}+i/2}
\prod_{j=1}^{M_3}\!\frac{u_{4,k}+u_{3,j}-i/2}{u_{4,k}+u_{3,j}+i/2}\!
\equiv \!{\rm BAE}_{4,k}.
\eeqa
We note that the magnons of the type $4$ can be split into two groups $u_{4,k}$ and $-u_{4,k}$ for $k=1,\dots,M_4/2$. Let us now prove some identities, and then proceed to write down the expression for the parity charge in this generic sector. Firstly, by considering the product of all auxiliary equations we get
\beq
\prod_{k=1}^{M_1}\!{\rm BAE}_{1,k}\prod_{k=1}^{M_2}\!{\rm BAE}_{2,k}\prod_{k=1}^{M_3}\!{\rm BAE}_{3,k} = \!\!\prod_{k=1}^{M_4/2}\!\[
\prod_{j=1}^{M_3}\frac{u_{4,k}-u_{3,j}+i/2}{u_{4,k}-u_{3,j}-i/2}
\prod_{j=1}^{M_3}\frac{u_{4,k}+u_{3,j}-i/2}{u_{4,k}+u_{3,j}+i/2}
\]
\eeq
which tell us that
\beq\la{u3u4}
\prod_{k=1}^{M_4/2}
\prod_{j=1}^{M_3}\frac{u_{4,k}-u_{3,j}+i/2}{u_{4,k}-u_{3,j}-i/2}
=
\prod_{k=1}^{M_4/2}
\prod_{j=1}^{M_3}\frac{u_{4,k}+u_{3,j}+i/2}{u_{4,k}+u_{3,j}-i/2}\;.
\eeq
Next computing the product of ${\rm BAE}_{4,k}$
over $k=1,\dots,M_4/2$ we get
\beqa\la{Pres0}
\!\prod_{k=1}^{M_4/2}\!\!
\[
\!\(\frac{u_{4,k}-i/2}{u_{4,k}+i/2}\)^{2L}
\prod_{j>k}^{M_4/2}\!\!\(\frac{u_{4,k}+u_{4,j}+i}{u_{4,k}+u_{4,j}-i}\)^2
\prod_{j=1}^{M_3}\frac{u_{4,k}-u_{3,i}-i/2}{u_{4,k}-u_{3,i}+i/2}
\frac{u_{4,k}+u_{3,i}-i/2}{u_{4,k}+u_{3,i}+i/2}
\]\!,
\eeqa
after using the identity \eq{u3u4} we notice that all terms are perfect squares of rational functions of Bethe roots.
Following our demonstrative example in the previous section, 
we know that the parity charge should be given be a square root of this expression. Indeed, we found the following expression to correctly reproduce the parity for the cases when the wave function is known explicitly\footnote{We have not tested the expression in the cases where the operators have spin, which may require an additional calibration by a sign factor depending on the quantum numbers.}:
\beqa\la{Pres}
{\mathbb P}=\prod_{k=1}^{M_4/2}
\[
-\(\frac{u_{4,k}-i/2}{u_{4,k}+i/2}\)^{L}
\prod_{j>k}^{M_4/2}\frac{u_{4,k}+u_{4,j}+i}{u_{4,k}+u_{4,j}-i}
\prod_{j=1}^{M_3}\frac{u_{4,k}-u_{3,i}-i/2}{u_{4,k}-u_{3,i}+i/2}\]\;.
\eeqa
Notice that this relation also has the following property: replacing any of $u_{4,k}$ by $-u_{4,k}$ does not change the value of ${\mathbb P}$, as the difference is proportional to the ${\rm BAE_{4,k}}$ which is $1$ when the roots satisfy the quantisation condition. This is a crucial consistency check, as it means that our expression is not sensitive to various ways of splitting the roots $u_{4,k}$ into two subsets of $M_4/2$ roots.
The expression (\ref{Pres}) allows us to know the parity eigenvalue of a state starting from the corresponding QSC solution.

\section{Constraining Linear Combinations of Squared OPE  Coefficients}\la{sec:ope_coefficients}
In addition to the bounds of $3$-point functions, presented in the previous section, we also analyse here a more general question, which is deducing bounds on arbitrary linear combinations of the squared OPE coefficients (assuming the coefficients are decaying sufficiently fast as we explain below). Such question is of course not new and appeared in various contexts before (see e.g.~\cite{Collier:2017shs,Paulos:2020zxx,Antunes:2021abs}). 

We will discuss our approach which is particularly fit for the $4$-point function itself, which can be considered as a linear combination of infinitely many OPE coefficients and it is an important physical quantity which we want to have under good numerical control. Potentially, the approach discussed here to treat the problem with semidefinite programming could be useful in some other situations too.

Another motivation for this study is the observation that whereas the OPE coefficients themselves could have very wide bounds, due to the degeneracies in the spectrum, certain linear combinations could be extremely narrow (as was already observed in \cite{Cavaglia:2021bnz}, see in particular the bounds of figure 4 there). Finding these linear combinations systematically could help to develop an efficient truncation scheme as an alternative to SDP.

\subsection{Modification to the crossing equations}
Let us denote the linear combination of the OPE coefficients as follows
\beq
T\equiv \sum_n C_n^2 H_{\Delta_n}
\eeq
where $H_{\Delta_n}$ are some weights, which, without reducing generality, we can assume to be real. In particular, in order to study the $4$-point itself, we can set these weights to be simply the relevant superconformal blocks, e.g.
\beq\label{eq:chooseHDelta}
H_{\Delta_n} = F_{\Delta_n}(x_0)\; .
\eeq
With this choice, we could directly relate the sum $T$ to the reduced correlator at cross ratio $x = x_0$, i.e. $T = f(x_0)-F_{\mathbb{I}}(x_0) -C^2_{\text{BPS} } F_{\mathcal{B}_2}(x_0)$ according to (\ref{eq:reducedOPE}). 
If we were instead interested to study the full-fledged amplitude $G(x_0)$ rather than its reduced cousin, we could just take
\beq\label{eq:chooseHDelta2}
H_{\Delta_n} = (2 x_0^{-1} - 1)F_{\Delta_n}(x_0) -\left(x_0^2 - x_0 +1\right)F_{\Delta_n}'(x_0).
\eeq
where we used the definition \eqref{pt4}.

Let us now proceed by considering a generic choice of these weights $H_{\Delta}$, which we assume to be a smooth functions of $\Delta$ (further requirements will be described below). In order to set up an optimisation procedure, we start from the crossing equation~\eq{eq:crossing0}
and, with a few manipulations, re-write it in the form
\begin{equation}\la{toopt}
\underbrace{\frac{\sum_{n=1}^{\infty} C^2_n H_{\Delta_n} }{ H_{\Delta_1}} }_{=T/H_{\Delta_1} }+\sum_{n>1} C^2_n\left(\frac{\mathcal{G}_{\Delta_n}(x)}{\mathcal{G}_{\Delta_1}(x)}-\frac{H_{\Delta_n}}{H_{\Delta_1}}\right) + \frac{\mathcal{G}_{\mathbb{I}}(x)+C_{\rm BPS}^2\mathcal{G}_{B_2}(x)}{\mathcal{G}_{\Delta_1}(x)}=0\; .
\end{equation}
Then we proceed in a standard way by trying to find a linear functional, acting on functions of the cross ratio $x$, with the following properties:
\begin{itemize}
    \item \texttt{Normalization}: $\alpha_\pm\(1\) = \pm 1$
    \item \texttt{Positivity}: $\alpha_\pm\(\sum_{n>1} C^2_n\left(\frac{\mathcal{G}_{\Delta_n}(x)}{\mathcal{G}_{\Delta_1}(x)}-\frac{H_{\Delta_n}}{H_{\Delta_1}}\right)\) \geq 0$ for $n=2,3,\dots$
    \item \texttt{Objective}: Maximizing $\alpha_\pm\left( \frac{\mathcal{G}_{\mathbb{I}}(x)+C_{\rm BPS}^2\mathcal{G}_{B_2}(x)}{\mathcal{G}_{\Delta_1}(x)}\right)$,
\end{itemize}
where in the first equation $\alpha_{\pm}()$ acts on the constant function of the cross ratio taking the value $1$ everywhere. Acting on the crossing equation with any functional satisfying the first two properties of the list above shows us that $\mp\alpha_\pm\left( \frac{\mathcal{G}_{\mathbb{I}}(x)+C_{\rm BPS}^2\mathcal{G}_{B_2}(x)}{\mathcal{G}_{\Delta_1}(x)}\right)$ gives an upper (lower) bound for $T/H_{\Delta_1}$, respectively. By finding a  functional satisfying the third optimisation requirement, we get an optimal bound. 

As in \cite{Cavaglia:2022qpg}, in practice  the positivity property is imposed for a number of the low-lying states of the spectrum (the ones we input from integrability), and after that it is imposed for all $\Delta \geq \Delta_{\text{gap}}$, where $\Delta_{\text{gap}}$ parametrises our ignorance of the high part of the spectrum, and we set it equal to the last level we know (in our case, $\Delta_{\text{gap}} = \Delta_{10}$). 

In order to guarantee this constraint on a semi-infinite range of $\Delta$ values with \texttt{SDPB}, one has to approximate the $\Delta$-dependence of the function  
\beq\label{eq:targetDelta}
\alpha_\pm\left(\frac{\mathcal{G}_{\Delta}(x)}{\mathcal{G}_{\Delta_1}(x)}-\frac{H_{\Delta}}{H_{\Delta_1}}\right)
\eeq
in the form of a polynomial with a positive prefactor. 
In the next section, we discuss in detail how to build this polynomial approximation for a functions of the form (\ref{eq:targetDelta}), which is a bit more subtle than for standard conformal blocks.\footnote{This could be useful for other contexts where one has sum rules with non-standard asymptotics, such as in the  integrated correlator constraints of \cite{Binder:2019jwn}, used for the bootstrap with linear programming in \cite{Chester:2021aun}. The approach we describe in section \ref{sec:twoscales}  could be useful for treating that problem with semidefinite programming without the need to discretize $\Delta$.} Before moving to this topic, let us briefly summarise what are the main parameters of the numerical implementation in \texttt{SDPB}, see \cite{Cavaglia:2022qpg} for more details:
\begin{itemize}
\item We restrict the search for linear functionals to a finite-dimensional space, built using the basis of functionals acting as
\beq\label{eq:basisalpha}
\alpha[h(x)]=\sum_{m=0}^{\Lambda/2} \alpha_m \left.\partial_x^{2m} h(x)\right|_{x=1/2}\;.
\eeq
The maximum order of derivatives $\Lambda$ considered is thus a truncation parameter for the problem. By the logic of these optimisation problems, the dependence of the bounds on $\Lambda$ is monotonic, i.e. they can only get better as $\Lambda$ is increased. 

\item A second integer parameter called $N_{\text{poles}}$ controls the order of the polynomial approximation discussed above. This parameter is chosen high enough that the effect on the value of the bounds is within an accepted tolerance. 
\end{itemize}

\subsection{Positivity at large $\Delta$: dealing with two exponential scales }\label{sec:twoscales}
When studying the problem of bounding the correlator at $x_0 \in [0,1/2)$, we find some subtlety which is absent in the more standard studies. To explain this point, let us recall the way \texttt{SDPB} handles the tail of the spectrum in the simple example of computing a bound for $C_1^2$ (i.e. the bootstrap problem we studied in \cite{Cavaglia:2021bnz}). In this case, the starting point is the crossing equation written in the form:
\beq\la{crossing2}
C^2_1 \underbrace{\mathcal{G}_{\Delta_1}(x)}_{\verb"norm"}+ 
\sum_{n>1} C^2_n \underbrace{\mathcal{G}_{\Delta_n}(x)}_{\verb"positive"} + 
\underbrace{\mathcal{G}_{\mathbb{I}}(x) +
C_{\rm BPS}^2\mathcal{G}_{B_2}(x)}_{\verb"objective"} = 0\;,
\eeq
and in order to get bounds on $C_1^2$, we have to find a functional $\alpha$, acting on functions of the cross ratio $x$, such that: 1) the action on the block $\mathcal{G}_{\Delta_1}$ -- i.e. the \verb"norm" term -- gives $\pm 1$, 2) the action on the terms under the sum gives a non-negative value, and 3) the action on the \verb"objective" is maximised. 
Given that we use the basis of functionals (\ref{eq:basisalpha}), 
 the problem is thus to find the set of $\left\{ \alpha_m \right\}$ for which the requirements above are satisfied. Note that for each $m$ the optimisation algorithm needs to know an infinite amount of quantities $\left.\partial_x^{2m}\mathcal{G}_{\Delta_n}(x)\right|_{x=1/2}$ for all $\Delta_n$, at least ideally. In practice, as explained above, only a few  $\Delta_n$ are known explicitly and therefore, on top of imposing constraints at those values individually, one imposes a stronger constraint that the functional produces non-negative values for all $\Delta$ above a certain $\Delta_{\rm gap}$, which is the last known dimension. 
 
 To implement this constraint one has to discretise the problem. The \texttt{SDPB} algorithm does this by using a polynomial approximation:
\beq\la{Gpaprox}
\left.\partial_x^{2m}\mathcal{G}_{\Delta}(x)\right|_{x=1/2} \simeq 
\verb"PREFACTOR"(\Delta)\;  P_m(\Delta-\Delta_{\rm gap})
\eeq
where  ${\verb"PREFACTOR"} (\Delta) $ is a positive (for $\Delta>\Delta_{\rm gap}$) common prefactor -- independent of $m$ and positive for $\Delta > \Delta_{\text{gap}}$--, which is mostly irrelevant for the optimisation problem, and $P_m(y)$. 
There are simple and well established ways of building such approximations, see e.g.~\cite{Chester:2021aun} for a review. After such polynomials are found, one can pass them to the \texttt{SDPB} package, which would find a solution to the optimisation problem constrained by the positivity condition
\beq
\sum_{m=0}^\Lambda \alpha_m P_m(y)\geq 0\;\;,\;\;y\geq 0 ,
\eeq
where usually $y=\Delta-\Delta_{\rm gap}$. The order of the polynomials is controlled by the parameter $N_{\text{poles}}$ discussed above. 

How can we also impose positivity of  the dimensions which are below $\Delta_{\rm gap}$ and which we know explicitly? For that we can use the option to pass to \texttt{SDPB} not just one polynomial per basis functional but several $P^a_m$, and impose
\beq
\sum_{m=0}^\Lambda \alpha_m P^a_m(y)\geq 0\;\;,\;\;y\geq 0\; ,
\eeq
where for $a=0$, $P^a_n(y)$ provides approximation of the \eq{Gpaprox}, and for $a=1,2,\dots$ those polynomials are simply constant functions given by $P^a_m(y)\equiv \left.\partial_x^{2m}\mathcal{G}_{\Delta_a}(x)\right|_{x=1/2}$, i.e. the values of the derivatives at fixed $\Delta=\Delta_a$, for all levels satisfying $\Delta_a < \Delta_{\text{gap}}$. 

To explain the main technical complication of applying the above method to our current setup, we have to look as the prefactor in \eq{Gpaprox}, which is taking care of the asymptotics of the conformal block at large $\Delta$. Conformal superblocks appearing in our theory decay exponentially as
\beq
F_{\Delta}(x) \sim \frac{(4 \rho(x) )^{\Delta} }{\Delta}\; ,
\eeq
where $\rho(x)=\frac{x}{\left(\sqrt{1-x}+1\right)^2}$. As we only need the values of the conformal block and its derivatives near 
$x=1/2$, the prefactor in \eq{Gpaprox} contains the exponentially decaying factor $(4\rho(1/2))^\Delta$, where $\rho(1/2)\simeq 0.171573$. The rest can be approximated well with a rational function with positive denominator. In our setup we have to approximate the following expression \eqref{toopt}
\beq\la{GH}
\left.
\d_x^{2m}
\(\frac{\mathcal{G}_{\Delta}(x)}{\mathcal{G}_{\Delta_1}(x)}-\frac{H_{\Delta}}{H_{\Delta_1}}\)\right|_{x=1/2} ,
\eeq
where $H_{\Delta}$ could, for example, be $F_{\Delta}(x_0)$ (where we can assume\footnote{Anyway, the remaining range of physical values $1/2<x_0<1$ can be reconstructed by crossing. Looking at analytic continuations of the cross ratio to other values, e.g. complex $x_0$ could be interesting but we leave this to future investigations.} $0<x_0<1/2$). We see that, while for $m>0$ the derivative just kills the second term in (\ref{GH}), for $m=0$ we now have two terms with different exponential asymptotics, with the second term decaying faster, 
\beq\la{GH2}
\left.
\d_x^{2m}
\(\frac{\mathcal{G}_{\Delta}(x)}{\mathcal{G}_{\Delta_1}(x)}-\frac{H_{\Delta}}{H_{\Delta_1}}\)\right|_{x=1/2} \simeq
\verb"PREFACTOR"(\Delta)\; \[ \tilde P_m(\Delta-\Delta_{\rm gap})+
\delta_{m,0}e^{-\alpha \Delta}Q(\Delta-\Delta_{\rm gap})
\] ,
\eeq
where $Q(y)$ is a polynomial, but we also have an exponential suppression with rate $\alpha = \log\left({\rho(1/2)}/{\rho(x_0)} \right) > 0$ in the current example.  

The challenge is to convert this second term in an effective way into a purely polynomial approximation on a semi-infinite interval to feed to \texttt{SDPB}. 

\paragraph{Polynomisation of the exponential.} So our task is to represent the expression in the square brackets in \eq{GH} as a polynomial in the semi-infinite interval $\Delta>\Delta_{\rm gap}$. One option would be to expand the exponential in a Taylor series, which may work well in some finite interval but will give an inadequate result for sufficiently large $\Delta$. So, we found the following method to work well. We split the semi-infinite interval into two: $[\Delta_{\rm gap},\Delta_{\rm up}]$ and $[\Delta_{\rm up},\infty]$, to be treated separately. The splitting point $\Delta_{\rm up}$ is chosen so that, in the second interval, the term $e^{-\alpha\Delta}$ can be neglected within our chosen target precision, and effectively to impose positivity in this range we can just neglect the contribution of $H_{\Delta}$ in (\ref{GH}). For the first interval we can, for example, use the Taylor expansion for the exponential or any other polynomial approximation. Then for the square bracket in \eq{GH} we get a polynomial of $\Delta$
\beq
[\dots]=\hat P_m(\Delta-\Delta_{\rm gap})\;.
\eeq
However, the problem is that \texttt{SDPB} is an algorithm to impose positivity on a semi-infinite, rather than finite, interval. To bring ourselves to that setting, we have to map the finite interval $\Delta \in [\Delta_{\rm gap},\Delta_{\rm up}]$ to $y \in [0, \infty]$, by defining a new variable $y$ in the following way\footnote{The same trick was used in \cite{Collier:2017shs} and goes back to David Simmons-Duffin. We are grateful to P.Kravchuk for pointing this out to us.}
\beq
y = \frac{\Delta_{\rm up} - \Delta_{\rm gap}}{\Delta - \Delta_{\rm gap}} - 1\;\;{\rm or}\;\;\Delta-\Delta_{\rm gap} = 
\frac{\Delta_{\rm up} - \Delta_{\rm gap}}{
y+1 }\;,
\eeq
such that in terms of $y$ we get
\beq
\hat P_m(\Delta-\Delta_{\rm gap})=
\frac{1}{
(y+1)^M }P^{\rm extra}_m(y)\;,
\eeq
where $M$ is the degree of the initial polynomial $P_m$.
We see that, up to an additional positive prefactor, we still get a set of polynomials whose linear combination has to be imposed to be non-negative for all $y>0$. This is precisely a semidefinite programming problem which can readily be tackled in \texttt{SDPB}.

We tried several options for approximating the exponential with the exponent. This includes a Taylor expansion in a point inside the interval and a Pad\`e approximation. But the most efficient way turns out to be the interpolation of the exponential using Chebyshev nodal points\footnote{We found that in some cases the approximation behaves slightly better if we also add the boundaries of the interval to the list of the points i.e. $i=1/2$ and $i=n+1/2$. }
\beq
d_i = \frac{\Delta_{\rm up}-\Delta_{\rm gap}}{2}\cos\left(\frac{(2i-1)\pi}{2n}\right) + \frac{\Delta_{\rm up}+\Delta_{\rm gap}}{2} \quad \text{for } i = 1,2,\ldots,n ,
\eeq
where $n$ is a parameter chosen to get the target precision, and then building the Lagrangian interpolation 
\beq
\exp(-\alpha \Delta)\simeq \sum_{i=1}^n\exp(-\alpha d_i)\prod_{j\neq i}\frac{\Delta-d_j}{d_i-d_j}\;.
\eeq
Using this method we managed to keep the maximal degree of the polynomials the same as for the standard problem of bounding $C_1^2$, while keeping the same precision of the approximation for all values of $\Delta$. We also compared the result with the Pad\'e approximation for some points (which does increase the degree of the polynomials and converges slower as a result) and fund no essential difference in the bounds within our precision.

\subsection{Including integrated correlator constraints}\label{sec:integratedbootstrap}

In order to include the integrated correlators constraint \eqref{eq:constr12} in the numerical bootstrap setup for linear combinations of the OPE coefficients, we follow the same approach introduced in \cite{Cavaglia:2022qpg}.
We rewrite the constraints (\ref{eq:constr12}) using the OPE decomposition \eqref{eq:reducedOPE} to obtain new linear relations for the OPE coefficients as follows
\beq\label{intcorr2}
\sum_{n}C_n^2 \int_{0}^{1/2} F_{\Delta_n}(x) \mu_a(x) dx+\texttt{RHS}_a(g)=0 \quad \text{with}\quad a=1,2
\eeq
where $\texttt{RHS}_a$ contains the contribution of the exact function $\mathcal{K}_a(g)$ in \eqref{eq:constr12} together with a contribution from the integral of the BPS conformal block appearing in \eqref{eq:reducedOPE} (see appendix \ref{app:app2}). The integration measures $\mu_a$  are simple rational functions of $x$. The two integrals appearing in \eqref{intcorr2} can be computed exactly and they are given in \ref{app:app2}.
 
In each of the two identities in (\ref{intcorr2}), we split the sum as
\beq
C_1^2\int_{0}^{1/2} F_{\Delta_1}(x)\mu_a(x) dx+
\sum_{n>1}C_n^2\int_{0}^{1/2}  F_{\Delta_n}(x)\mu_a(x)dx + \texttt{RHS}_a(g)=0\;.
\eeq
Proceeding like the previous sections, we add and subtract a piece which highlights the quantity $T = \sum_n H_{\Delta_n}$ we are interested in.  This yields the following exact rewriting
\begin{equation}
\frac{T}{H_{\Delta_1}} +\sum_{n>1} C^2_n\left(\frac{\int_{0}^{1/2}  F_{\Delta_n}(x)\mu_a(x)dx}{\int_{0}^{1/2} F_{\Delta_1}(x)\mu_a(x)dx}-\frac{H_{\Delta_n}}{H_{\Delta_1}}\right) + A_a(g) = 0
,
\;\;a=1,2 ,\end{equation}
with $A_a(g) \equiv  \texttt{RHS}_a(g)/{\int_{0}^{1/2} F_{\Delta_1}(x)\mu_a(x) dx} $. 

Then, following \cite{Cavaglia:2022qpg} we can simply consider arbitrary  linear combinations of these two identities with the $1 +  \Lambda/2$ nontrivial relations obtained by taking $m=0,1,\dots,\Lambda/2$ non-trivial derivatives $\partial^{2m}_x$ of (\ref{toopt}) in the cross ratio at $x=1/2$. 
We can setup an optimisation problem for \texttt{SDPB} acting in the space of coefficients of this linear combination. For completeness we give the full details below.

The functionals are represented as co-vectors with $\Lambda/2+3$ real components:
\beq
\vec{\alpha} \equiv \left(\alpha_0, \dots \alpha_{\Lambda/2} | \alpha_{1}^{\texttt{Int}}, \alpha_{2}^{\texttt{Int}} \right) ,
\eeq
where $\Lambda$ is the cutoff on the number of  derivatives. 
For each $\Delta$, we define a vector $V_{\Delta} \in \mathbb{R}^{\Lambda/2+3}$,
 \beq
\vec V_{\Delta} \equiv\left. \left(  \rho_{\Delta}(x) ,   \partial_x^{2} \rho_{\Delta}(x) , \dots,  \partial_x^{ \Lambda} \rho_{\Delta}(x) \right)\right|_{x=1/2} \oplus \left( R_{\Delta,1}, \; R_{\Delta, 2}\right) ,
 \eeq
 with 
 \beq\label{eq:defnewblocks}
\rho_{\Delta}(x) \equiv \frac{\mathcal{G}_{\Delta}(x)}{\mathcal{G}_{\Delta_1}(x)}-\frac{H_{\Delta}}{H_{\Delta_1}}, \;\;\; R_{\Delta,a}\equiv \frac{\int_{0}^{1/2}  F_{\Delta}(x)d\mu_a}{\int_{0}^{1/2} F_{\Delta_1}(x)d\mu_a}-\frac{H_{\Delta}}{H_{\Delta_1}} , \; a = 1,2 ,
 \eeq
so that the action of the functional on the vector defines a function of $\Delta$, i.e. $\vec{\alpha} \cdot V_{\Delta}$. 

 The optimisation problem is defined as follows. Find $\vec{\alpha}$ such that
\begin{enumerate}[1)]
\item \texttt{Normalization}:
\beq
\vec{\alpha}_{\pm} \cdot  \vec{\texttt{norm}}= \pm 1,
\eeq
with
\beq
\vec{\texttt{norm}} =  (  1, \underbrace{0,\dots, 0 }_{=\vec{0} \in \mathbb{R}^{\Lambda/2} } ) \oplus \left( 1, \; 1\right) .
\eeq
\item \texttt{Positivity}: $\vec{\alpha}_{\pm} \cdot \vec{V}_{\Delta} \geq 0$ for $\Delta \geq \Delta_{\text{gap}}$, as well as for $\Delta = \Delta_n$ where $\Delta_n$ are all states of the spectrum with $\Delta_n < \Delta_{\text{gap}}$. As usual, we choose $\Delta_{\text{gap}}$ to coincide with an excited state in the spectrum, typically $\Delta_{\text{gap} }= \Delta_{10}$ in our studies. 
\item \texttt{Objective}: the quantity $\vec{\alpha}_{\pm}\cdot \vec{\texttt{obj}}$ is maximised, with
\beq
\vec{\texttt{obj}} =\left.  \left( \partial_x^0, \partial_x^2 \dots, \partial_x^{\Lambda} \right)\left[ \left( \frac{\mathcal{G}_{\mathbb{I}}(x)+C_{\rm BPS}^2\mathcal{G}_{B_2}(x)}{\mathcal{G}_{\Delta_1}(x)}\right)\right] \right|_{x = \frac{1}{2}} \; \oplus \; \left( A_1(g), A_2(g)\right). 
\eeq
\end{enumerate}
By construction, the latter quantity will give  us an upper/lower bound:
\beq
\vec{\alpha}_{-}\cdot \vec{\texttt{obj}} < T/H_{\Delta_1} < \vec{\alpha}_{+}\cdot \vec{\texttt{obj}}. 
\eeq
To enforce positivity we should produce a polynomial-type approximation for the last two components of $V_{\Delta}$, i.e. the ones related to the integrated correlator constraints. It was noticed in \cite{Cavaglia:2022qpg} that these integrals have the same exponential behaviour at large $\Delta$ as the derivatives of the blocks. Moreover, we see in (\ref{eq:defnewblocks}) that we are subtracting the exact same term $-H_{\Delta}/H_{\Delta_1}$ to  all components of the vector $V_{\Delta}$. Thus, we can  build the polynomial-type approximation of the two functions $R_{a, \Delta}$, $a = 1,2$ in full parallel with the other components of $V_{\Delta}$, following the steps discussed in section \ref{sec:twoscales}. 

\section{Results for Bounds on the Four-Point Function}\label{sec:4pt}
In this section we present our numerical results for the 4-point function, which are reported in appendix \ref{app:app3} and in an attached file. We computed bounds  for the full correlator $G(x)$ of four tilt operators $\Phi_M$, all with the same index for definiteness. We also obtained bounds directly for the reduced correlator $f(x)$, in terms of which one can easily obtain the result for all other polarisations of the four  external BPS multiplets~\cite{Liendo:2018ukf}. 

\paragraph{Numerical implementation. } 

To bound $G(x)$, we implement the algorithm choosing  the weight function $H_{\Delta_n}$ defined in (\ref{eq:chooseHDelta2}), which allows us to reconstruct bounds for the contribution of non-protected operators to the correlator $G(x)$, i.e. with this choice we bound the quantity
\beq\label{T1}
\frac{T}{H_{\Delta_1}} =  \frac{G_{\text{NBPS}}(x)}{H_{\Delta_1}} \equiv \frac{G(x) - 1 - C_{\rm BPS}^2\left( x^2 + \frac{2-x}{x} F_{\mathcal{B}_2}(x) + (-1 + x -x^2)F_{\mathcal{B}_2}'(x)  \right)}{ \frac{2-x}{x} F_{\Delta_1}(x) + (-1 + x -x^2)F_{\Delta_1}'(x)}
\eeq
for generic $x \in [0,1/2]$, from which we can easily reconstruct the value of $G(x)$ and also extend it to the full interval $x \in [0,1]$ using \eqref{crossG}. To bound instead directly the reduced correlator $f(x)$, we chose the weight function $H_{\Delta_n}$ as in (\ref{eq:chooseHDelta}). Then, the quantity of interest is the following
\beq\label{T2}
\frac{T}{H_{\Delta_1}} =  \frac{f_{\text{NBPS}}(x) }{H_{\Delta_1}}\equiv \frac{f(x) - x - C_{\rm BPS}^2F_{\mathcal{B}_2}(x)}{F_{\Delta_1}(x)}
\eeq
where, in both \eqref{T1} and \eqref{T2}, we used the definitions \eqref{pt4} and \eqref{eq:reducedOPE} and the explicit value of $F_{\mathbb{I}}$ given in Appendix \ref{app:app1} together with $F_{\mathcal{B}_2}$.

Our main results were obtained by incorporating the two integrated correlator identities and the parity selection rule throughout in the algorithm. In particular, we input the knowledge of the dimensions of the 10 lowest-lying non-protected states (of which one drops from the analysis because of parity considerations, as described in the previous section). 

The data discussed below are obtained using a truncation to $\Lambda = 90$ derivatives\footnote{We also collected data with $\Lambda = 30$ and $\Lambda = 60$. Comparing $\Lambda = 30$ and $\Lambda = 90$, we see a gain in precision by roughly one order of magnitude at weak coupling, and more than two orders at strong coupling for $g\sim 4$.}, and a polynomial approximation of the blocks involving polynomials of degree $N_{\text{poles}} = 40$, both for the interval $\Delta \in [ \Delta_{\text{gap
}}, \Delta_{\text{up}}]$  and for the separate approximation for $\Delta \in [ \Delta_{\text{up}}, +\infty]$. The value of the splitting point $\Delta_{\text{up}}$ is chosen in such a way that we keep the same precision despite dropping the exponentially subleading term in the second interval.
Thus, the value of $\Delta_{\text{up}}$ used to collect our data varies with the value of $x_0$, and in principle becomes very big when $x_0$ approaches 1/2. However, given our parameters, we noticed that setting a cutoff for $\Delta_{\text{up}}$ at $\sim 100$ does not affect the \texttt{SDPB} output within our target precision. 

The spectral data we input have $\sim 20$ digits precision. We estimate that the impact of this source of error, together with the truncation errors involved in the approximation of the blocks, affect the results by an error that is negligible as it is some orders of magnitude smaller than the width of the numerical bootstrap bounds.

\begin{figure}
   \centering
    \includegraphics[width=\dimexpr 0.49\columnwidth\relax]{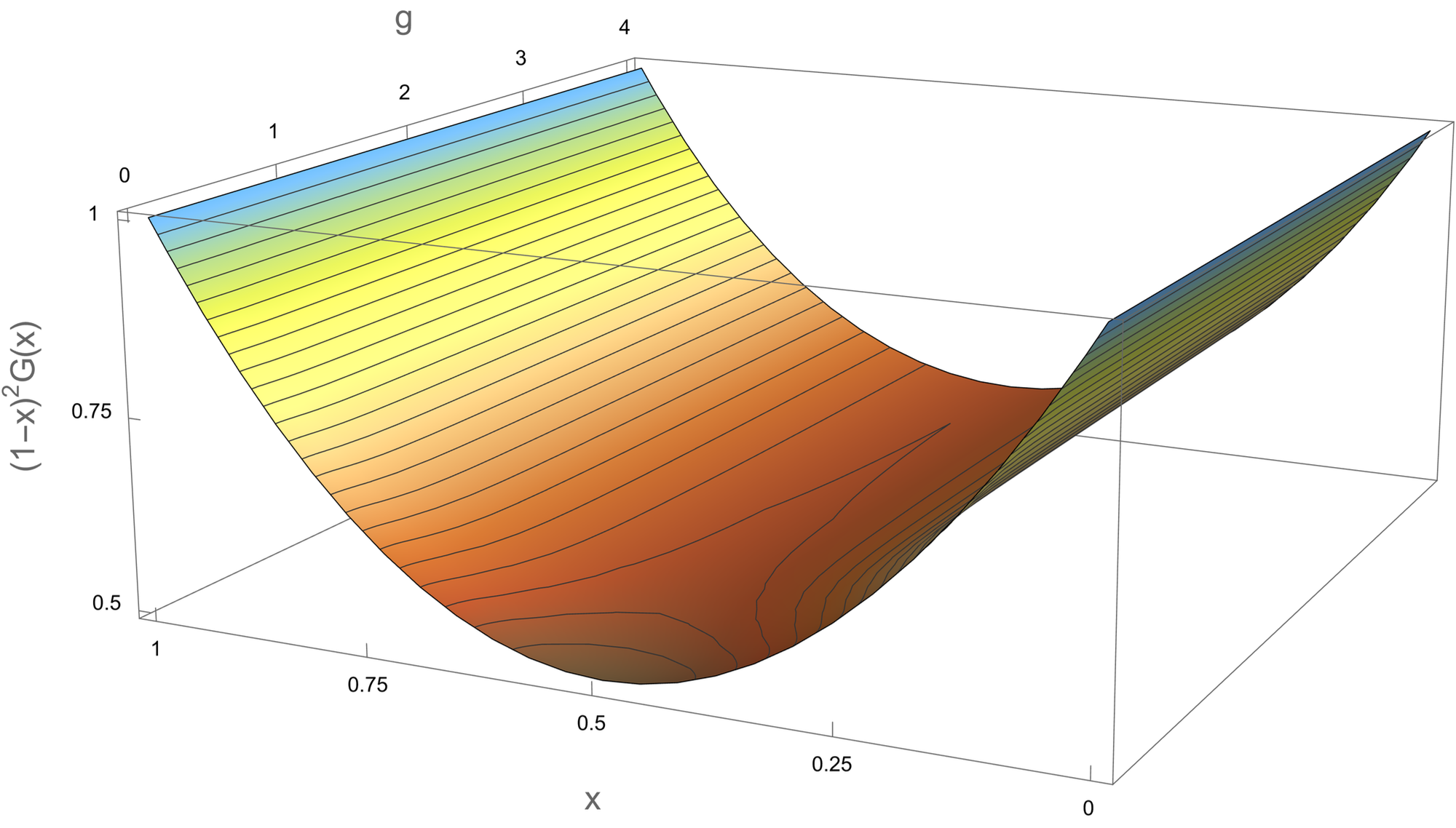}
    \includegraphics[width=\dimexpr 0.49\columnwidth\relax]{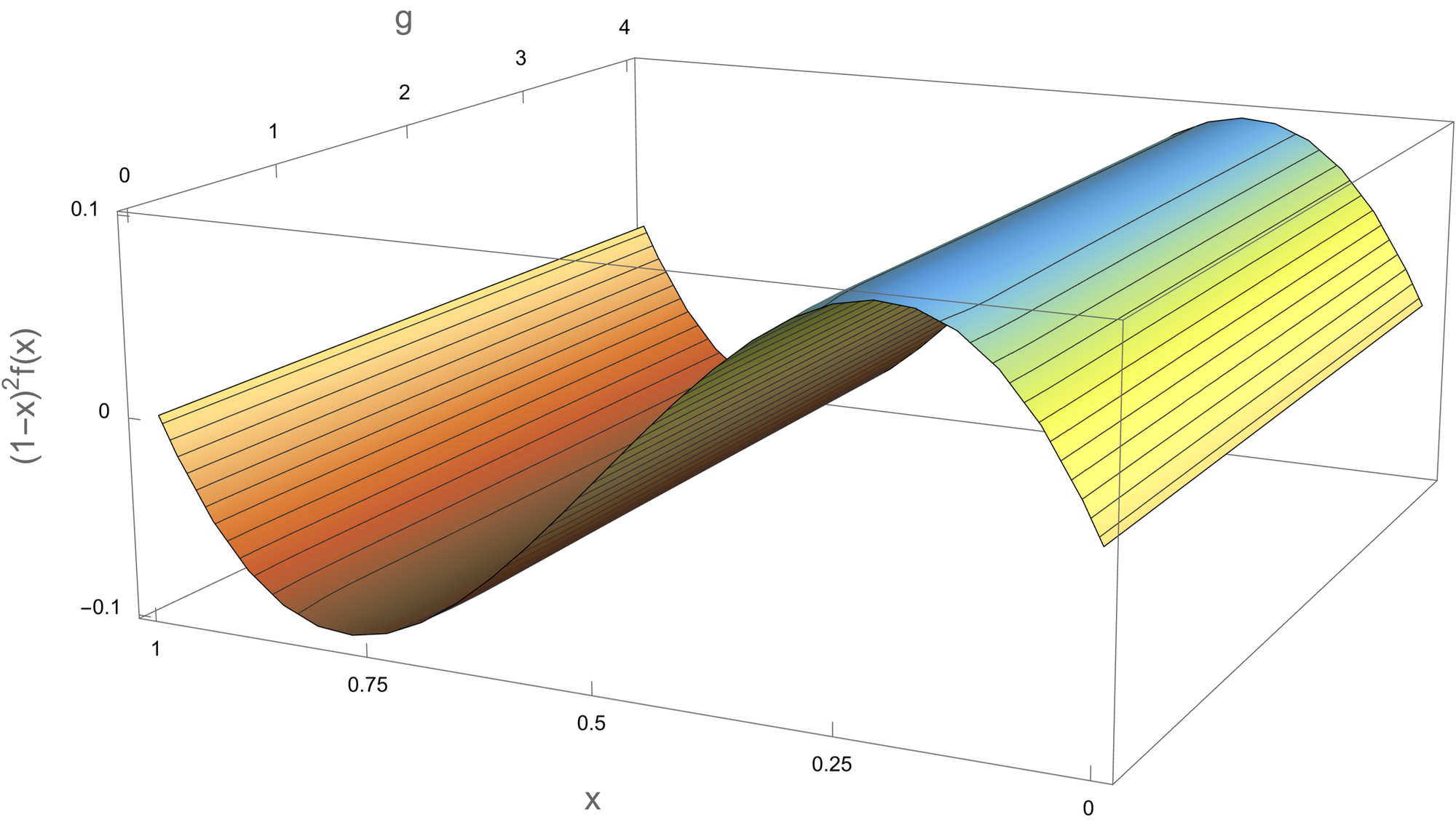}
   \caption{Upper and lower bounds for $\tilde{\mathcal{G}}(x) = (1-x)^2 G(x)$ (left), and of $(1-x)^2 f(x)$ (right) over values of the coupling and cross ratio $g \in [0,4]$, $x\in [0,1]$.
   The bounds are very narrow (see fig. \ref{fig:1b}), and their width is invisible on the scale of the plots (in fact, the thickness of the lines used in the plots is far wider).
   }
    \label{fig:1a}
\end{figure}

The results obtained in this way for the two functions $G(x)$ and $f(x)$, multiplied by a factor $(1 - x)^2$ to make them crossing (anti)-symmetric, are shown in the left and right panels of figure \ref{fig:1a}, over the range of cross ratio and coupling constant.  Notice that the lower and upper bounds are so narrow that we get what look like exact plots! More detailed pictures of the results are presented in figures \ref{fig:Gdetail} and \ref{fig:fdetail}. 

\begin{figure}

   \centering
    \includegraphics[width=\dimexpr 0.49\columnwidth\relax]{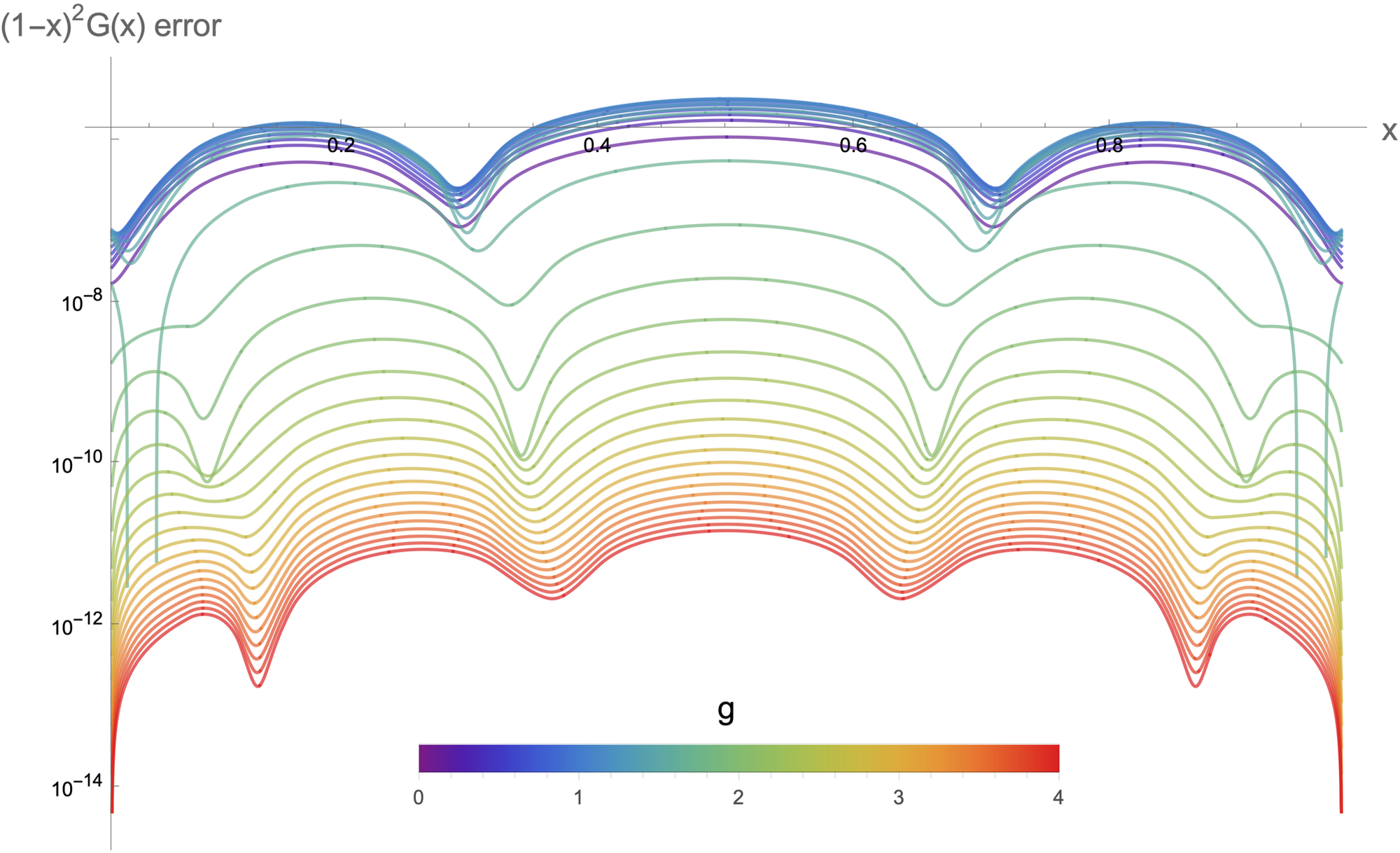}    
    \includegraphics[width=\dimexpr 0.49\columnwidth\relax]{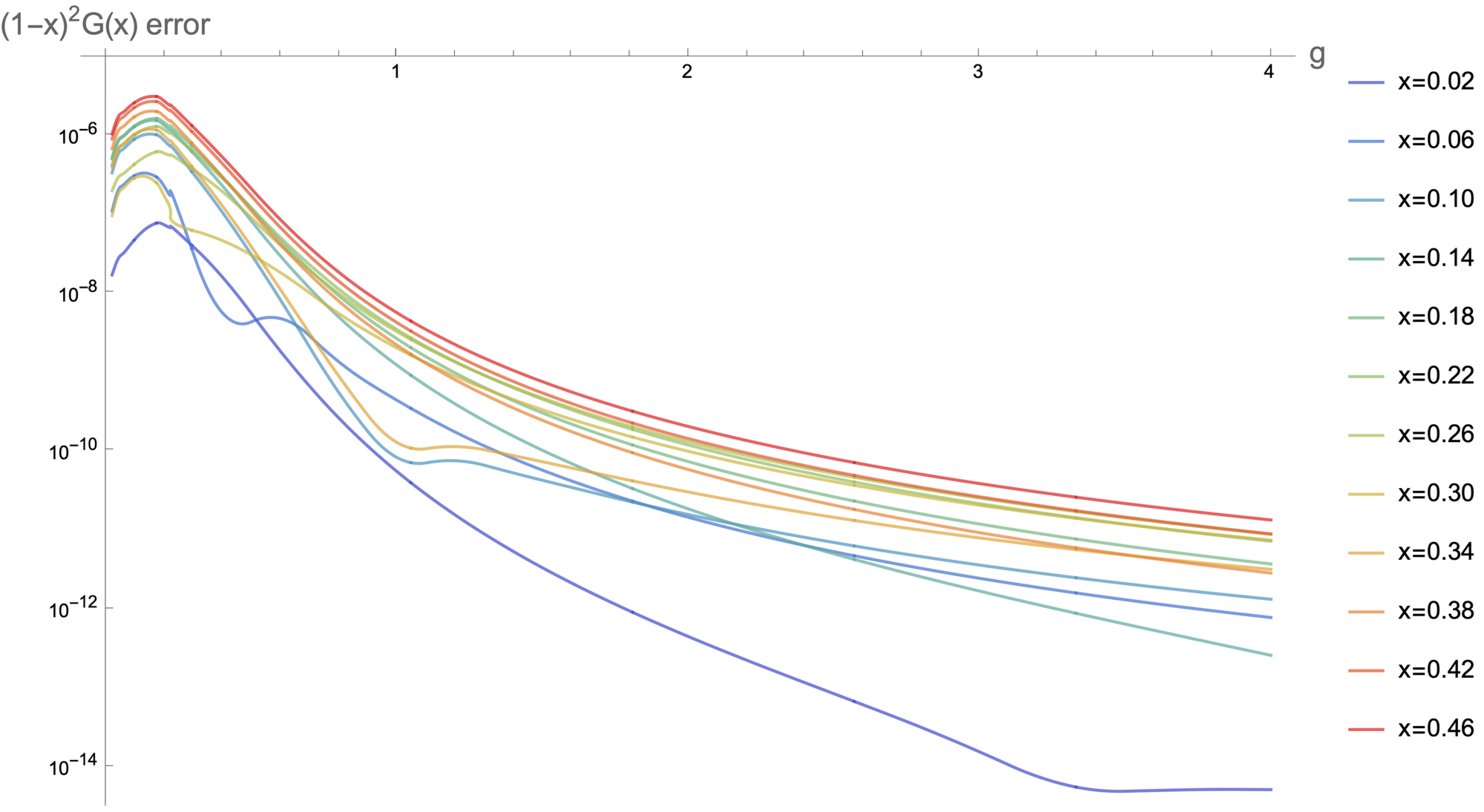}
    
    \includegraphics[width=\dimexpr 0.49\columnwidth\relax]{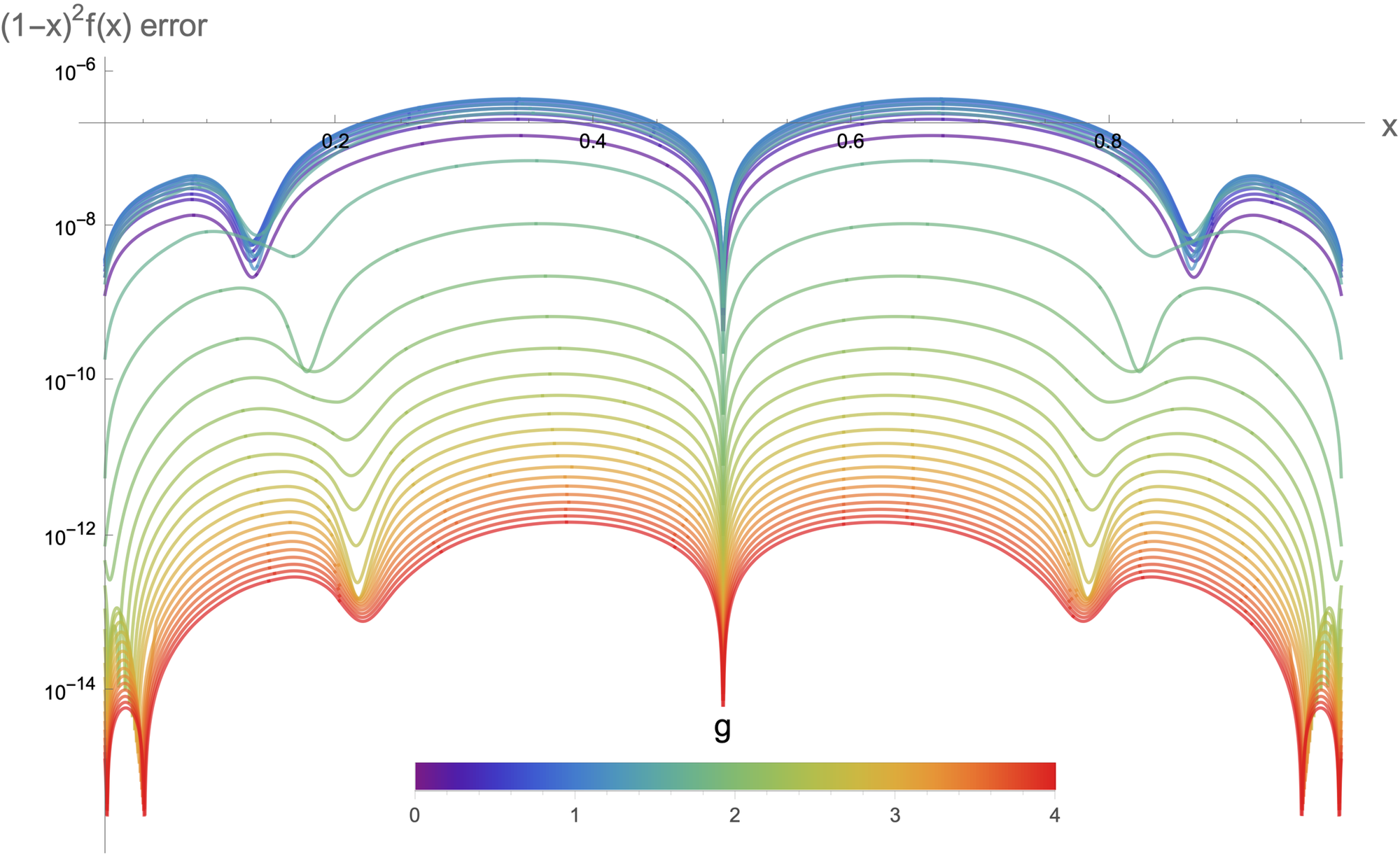}
     \includegraphics[width=\dimexpr 0.49\columnwidth\relax]{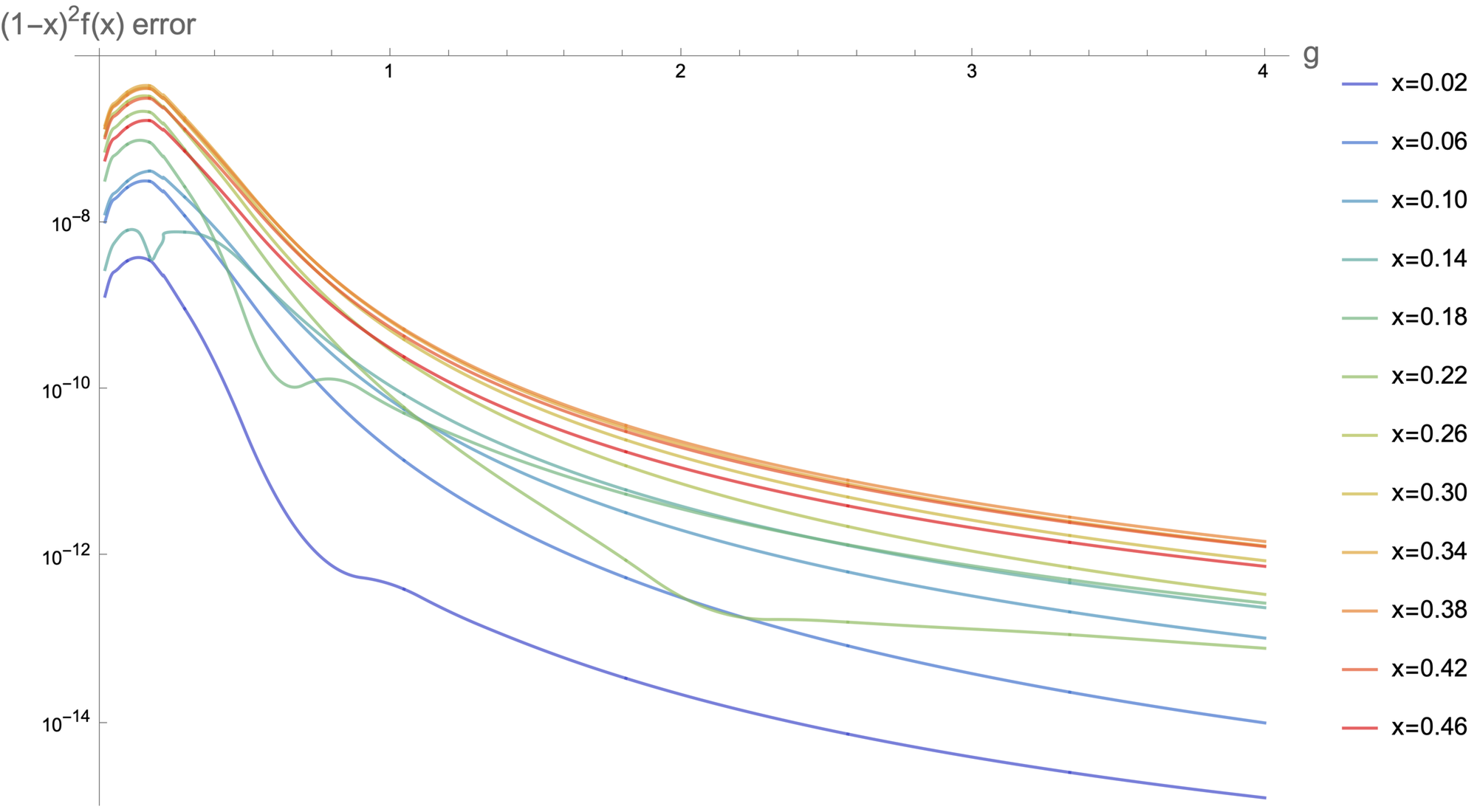}
    \caption{The error (evaluated as half of the width of the bounds) for the 4-point function $G(x)$ (first line) and reduced amplitude $f(x)$, (second line).
    The error is plotted against the coupling and against the cross ratio in the left vs right panels. 
    It is apparent how the error drops dramatically with increasing value of the coupling, as well as close to the edges of the interval $[0,1]$ in the cross ratio. 
    }
    \label{fig:1b}
\end{figure}

\paragraph{Bounds width.}
The error, evaluated as half the width of the bounds, is depicted in the panels of figure \ref{fig:1b}, for the two crossing-symmetrised quantities $(1-x)^2 G(x)$ and $(1-x)^2 f(x)$. 

As already observed for bounds for OPE coefficients, the precision rapidly improves with stronger coupling giving e.g., for $(1-x)^2 G(x)$, bounds with 10-11 digits precision for the whole range of the cross ratio! Even at weak coupling, we get roughly at least $6$ digits precision. 

We remark that with the previous method of \cite{Cavaglia:2022qpg} we obtained this kind of precision \emph{only for} the leading OPE coefficient $C_1^2$ at very weak and very strong coupling, while the precision drops rather significantly for excited states (around 5 digits). Moreover, this refers to the collection of data with $\Lambda=140$ while here we consider $\Lambda=90$. 

Thus, taking those bounds and trying to reconstruct the 4-point by plugging them into the OPE, we would not have got very precise results. 

\paragraph{Comments on the 4-point function.}

\begin{figure}
    \centering
    \includegraphics[width=\columnwidth]{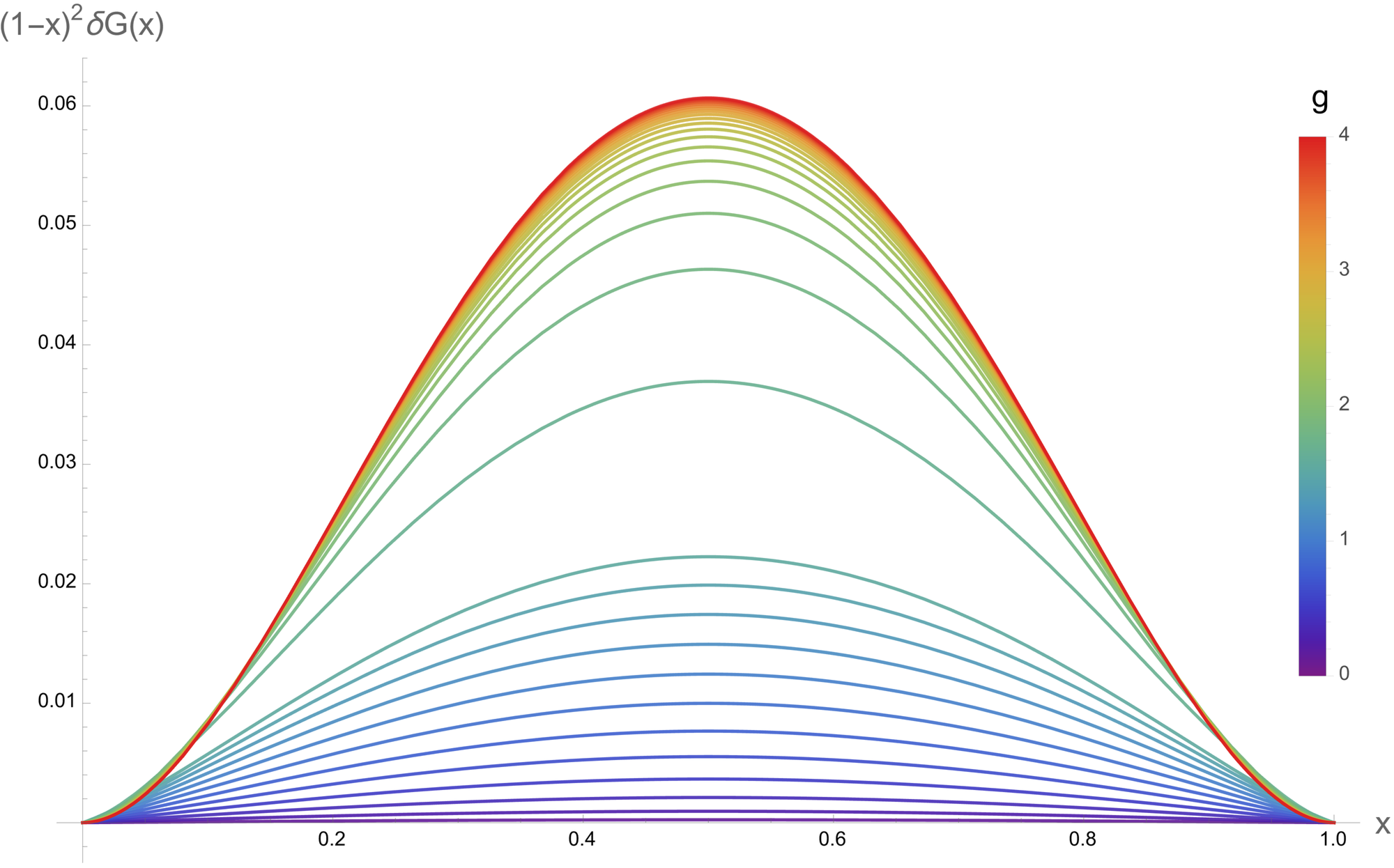}
    \caption{A detailed depiction of the 4-point function. Here we plot the bounds for $(1-x)^2 \delta G(x) \equiv (1-x)^2 G(x) - (1-x)^2 G^{(0)}(x)$, where $(1-x)^2 G^{(0)}(x) = 1 - 2 x + 2 x^2$ is the zero coupling value. This subtraction magnifies  the variation in the coupling, which is represented by different colours. Again, the bounds are very narrow and the lines used to make the plot visible are actually thicker than them.}
    \label{fig:Gdetail}
\end{figure}

\begin{figure}

   \centering
    \includegraphics[width=\dimexpr \columnwidth\relax]{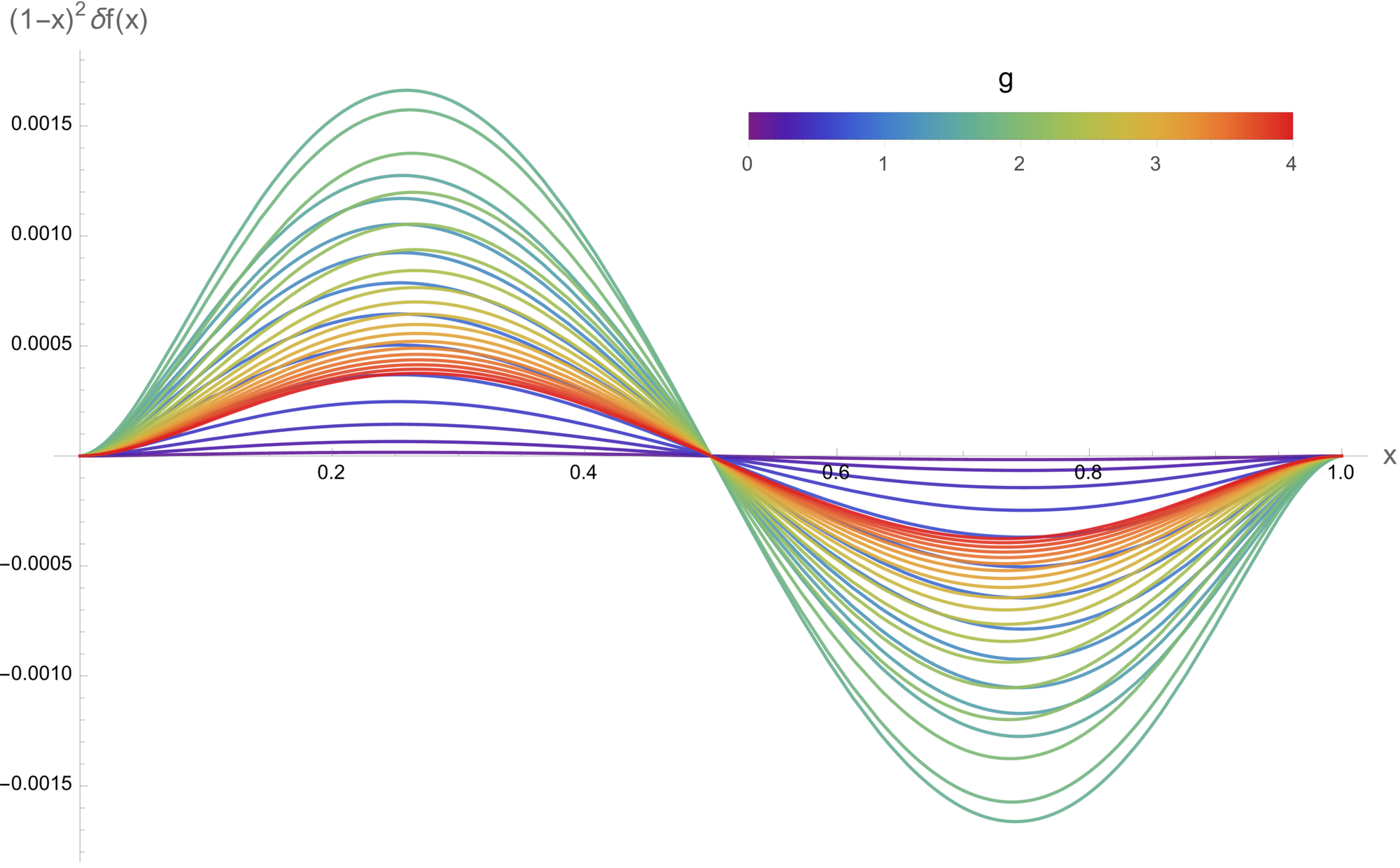}
    \caption{A detailed view of the result for the reduced correlator. Here we plot the function $(1-x)^2 \delta f(x) = (1-x)^2 (f(x) - \frac{2 x^2 - x}{x-1} )$, over the cross ratio, representing different values of $g$ by different colours.  The subtraction of the zero coupling value in the definition of $\delta f(x)$ is useful to magnify the effects of varying $g$.
    }
    \label{fig:fdetail}
\end{figure}

\begin{figure}
    \centering
    \includegraphics[trim={0 4.2cm 4cm 0},clip,width=\columnwidth]{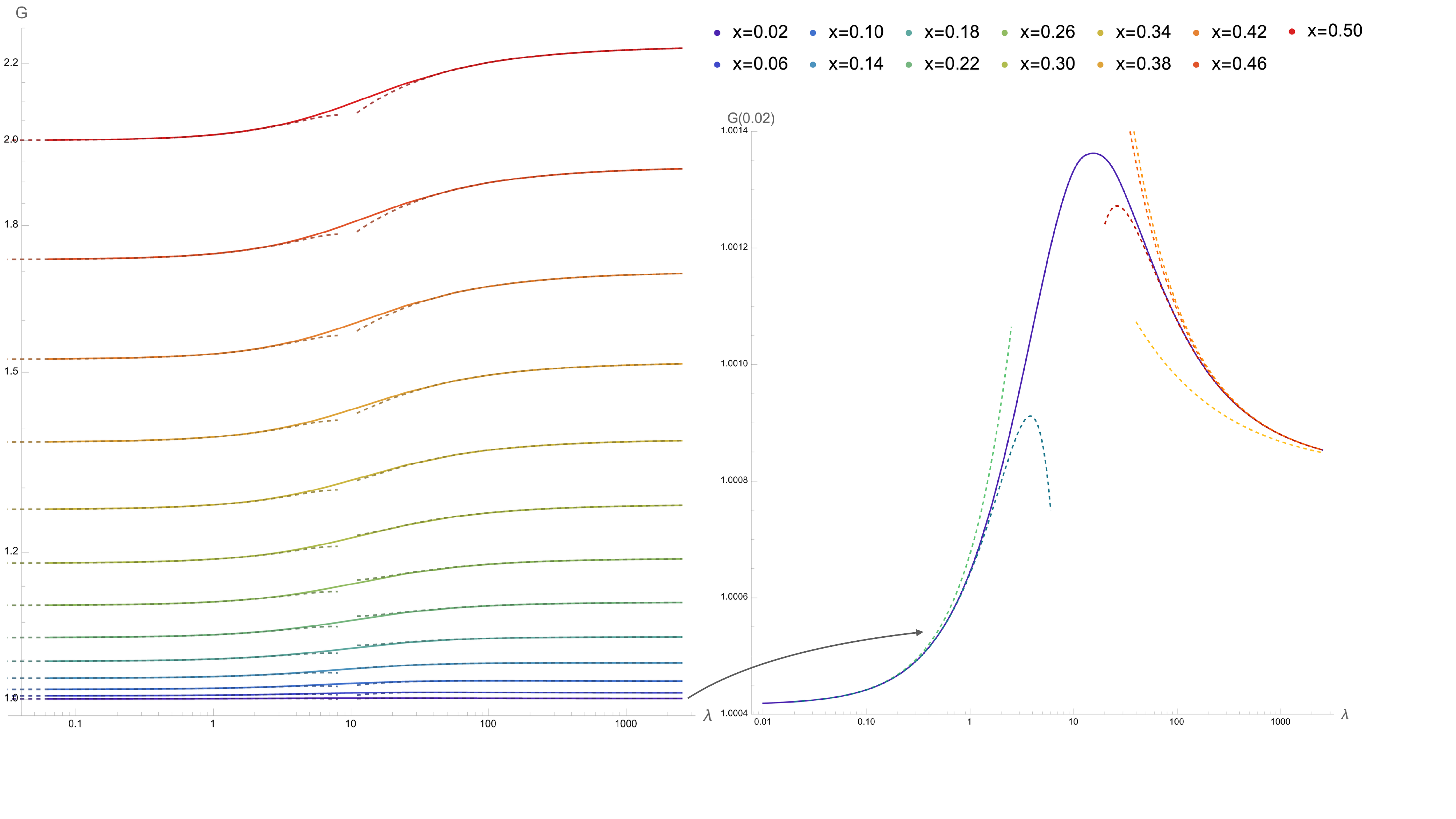}
    \caption{On the left: $G(x)$ as a function of the coupling $\lambda$ in Log scale for fixed $x$, together with weak and strong coupling predictions \cite{Ferrero:2021bsb,Cavaglia:2022qpg}. On the right: the zoom on $G(x)$ for $x=0.02$ showing the non-monotonicity for small values of $x$. The function $G$ start to be monotonic in $\lambda$ from a value of $x$ close to 0.15. Dashed lines corresponds to progressively precise weak (\textcolor{color1}{$\lambda$} and \textcolor{color2}{$\lambda^2$}) and strong coupling (\textcolor{color3}{$1/\lambda^{1/2}$}, \textcolor{color4}{$1/\lambda$} ,\textcolor{color5}{$1/\lambda^{3/2}$} and \textcolor{color6}{$1/\lambda^{2}$}) predictions.}
    \label{fig:compare1}
\end{figure}

Let us make some comments on the shape of the 4-point function, more precisely of the natural   crossing-symmetric combination 
\beq
\tilde{\mathcal{G}}(x)\equiv G(x) (1-x)^2 ,\;\;\; \tilde{\mathcal{G}}(x) = \tilde{\mathcal{G}}(1-x).
\eeq
From our bounds we can infer that, at generic finite coupling $g >0$ and $x \in [0,1]$, $\tilde{\mathcal{G}}(x)$ takes values between $1$ and $1/2$. As a welcome sanity check, we see that our lower bounds always satisfy the strict inequality
\beq\label{eq:boundsFB}
\tilde{\mathcal{G}}_F(x) < \tilde{\mathcal{G}}(x) ,
\eeq
where $\tilde{\mathcal{G}}_{F}(x) = \left( - 1 + x^{-2} + (1-x)^{-2}\right) x^2 (1-x)^2$ corresponds to the 4-point function in a fermionic Generalized Free Field theory (GFF) with four external fields of dimension $1$. This is a universal bound derived in \cite{Paulos:2020zxx} to be true for any unitary CFT$_1$. Our theory explores part of this allowed region as the coupling is varied. It appears that the theory satisfies a stricter lower bound, which is saturated only at $g = 0$:
\beq
\tilde{\mathcal{G}}(x) \geq \tilde{\mathcal{G}}^{(0)}(x) > \tilde{\mathcal{G}}_F(x)\;,
\eeq
where $\tilde{\mathcal{G}}^{(0)}(x)$ is the zero coupling value, 
\beq
\left. \tilde{\mathcal{G}}^{(0)}(x) \equiv \tilde{\mathcal{G}}(x)\right|_{g = 0} = 1 - 2 x + 2 x^2 > \tilde{\mathcal{G}}_F(x), \;\; x \in [0,1] .
\eeq
At strong coupling, instead, the theory approaches asymptotically the bosonic GFF value,
\beq
\lim_{g\rightarrow +\infty} {\tilde{\mathcal{G}}}(x) =  {\tilde{\mathcal{G}}}_B(x), \;\; x \in [0,1],
\eeq
where $\tilde{\mathcal{G}}_{B}(x) = \left( + 1 + x^{-2} + (1-x)^{-2}\right) x^2 (1-x)^2$.
However, this is \emph{not} an upper bound, since there are values of the cross ratio, close to the boundaries of the interval $[0,1]$, where the correlator exceeds this value for some values of the coupling. This is already clear from the strong coupling perturbative results of  \cite{Ferrero:2021bsb}, and we exhibit on our numerical data in  figure \ref{fig:compare1}.\footnote{Notice that in \cite{Paulos:2020zxx} an upper bound equal to $\tilde{\mathcal{G}}_{B}(x)$ was deduced for theories that satisfy a certain condition on the spectrum that our theory does not obey: having a gap given by twice the dimension of the external operator. In our case, this condition is not met since the operator $\Phi_{6}$ has a dimension interpolating between $1$ and $2$, see fig. \ref{fig:spectrum10}. The fact that our theory exceeds the bosonic GFF values shows that dropping this gap assumption may indeed lead to a violation of the upper bound.}

\begin{figure}
    \centering
    \includegraphics[trim={0 4.2cm 3.5cm 0},clip,width=\columnwidth]{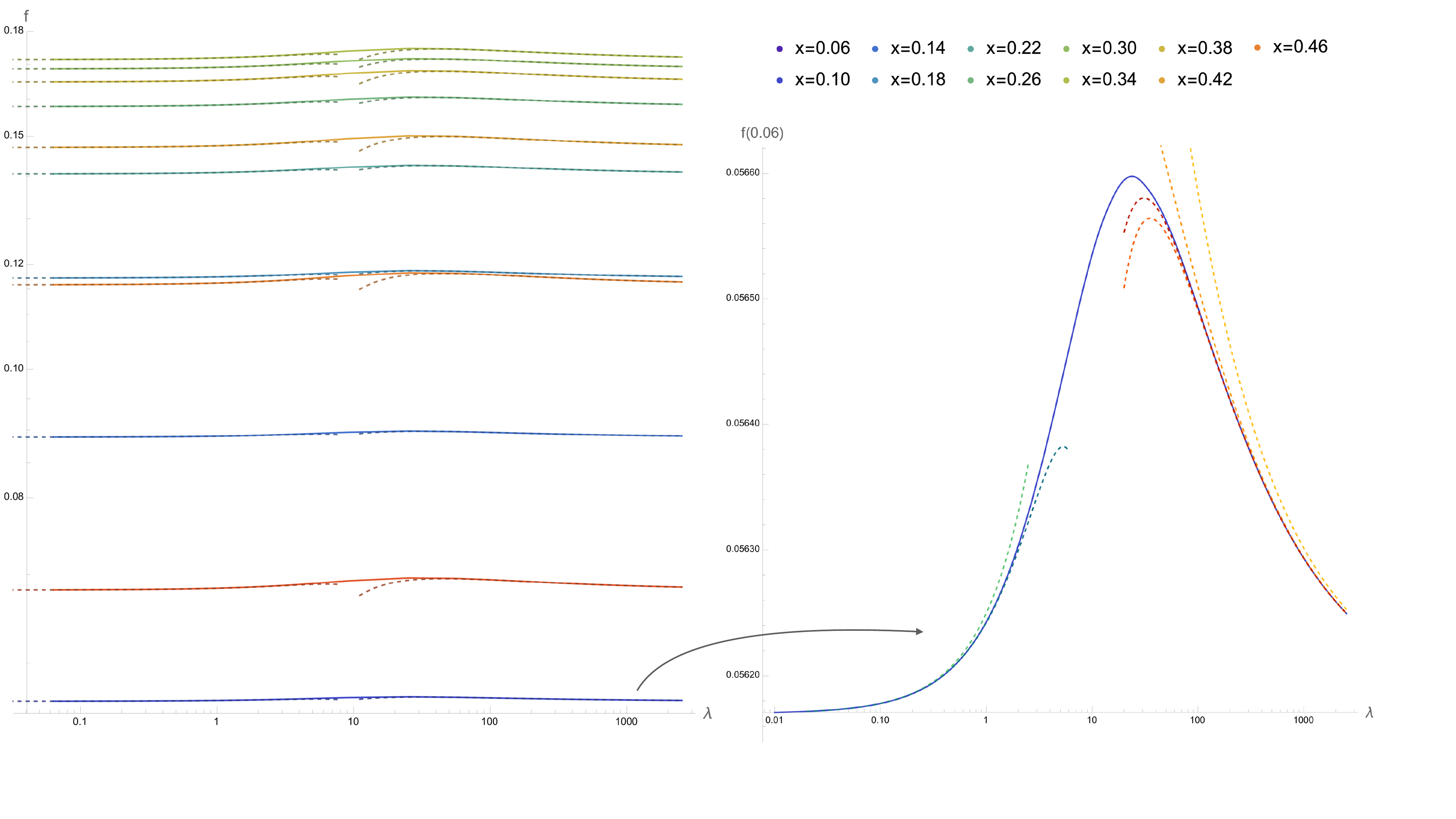}
    \caption{On the left: $f(x)$ as a function of the coupling $\lambda$ in Log scale for fixed $x$, together with weak and strong coupling predictions \cite{Ferrero:2021bsb,Cavaglia:2022qpg}. On the right: the zoom on $f(x)$ for $x=0.06$ showing the non-monotonicity in $\lambda$. Notice that this property is true for any value of $x$. Dashed lines corresponds to progressively precise weak (\textcolor{color1}{$\lambda$} and \textcolor{color2}{$\lambda^2$}) and strong coupling (\textcolor{color3}{$1/\lambda^{1/2}$}, \textcolor{color4}{$1/\lambda$} ,\textcolor{color5}{$1/\lambda^{3/2}$} and \textcolor{color6}{$1/\lambda^{2}$}) predictions.}
    \label{fig:compare2}
\end{figure}

\paragraph{Comparison with perturbative  results.}
Figure \ref{fig:compare1} and \ref{fig:compare2} shows how our data interpolate nicely between the known perturbative results for $G(x)$ and $f(x)$~\cite{Kiryu:2018phb,Cavaglia:2022qpg,Ferrero:2021bsb}, i.e. 5 orders at strong, which can be found in the attached material to \cite{Ferrero:2021bsb},  and 3 orders at weak coupling, which can be found in \cite{Cavaglia:2022qpg}. 

Notice that, while the dependence on $g$ may appear always monotonic, there are values of the cross ratio where this is definitely not the case. This is visible clearly in the right side of figure \ref{fig:compare1}, where we zoom on the curve with cross ratio $x = 0.02$.

\section{Discussion}\la{sec:discussion}

In this paper we considered the implications of the parity symmetry on the bounds for the structure constants and also obtained new bounds directly for the 4-point function at the physical values of cross ratio $x\in [0,1]$. We see that, while there is still no analytical solution for this quantity, we got a surprisingly accurate picture for a correlation function at finite 't Hooft coupling and operator positions.

While our results are numerical and in a very particular setup, they  are feeding towards a more conceptual discussion, initiated in~\cite{Cavaglia:2021bnz}, on how much the data from integrability for the spectrum of integrable AdS/CFT dualities, combined  with the constraints  of  crossing symmetry, determine the rest of the conformal data and higher point correlation functions.
The narrow bounds obtained in this paper for a 4-point function give further support to this idea. 

Among the interesting questions for the future, it would be intriguing to see how far one can continue the cross ratio into the complex plane of $x$ while keeping the bounds reasonably narrow. Is it possible to get insights into the region describing Lorentzian kinematics with this method? It would be interesting if one could approach the OTOC regime, studied in a double scaling limit at strong coupling in \cite{Giombi:2022pas}. 

While this simple single-correlator problem provides tight bounds for some quantities, going forward with the Bootstrability program it seems crucial to incorporate more correlators in the setup. We found that for protected external operators one can get a further reduction of the free parameters using parity symmetry as discussed in section \ref{sec:complex}, and results for this setup will be published in \cite{MultiC}. 
Furthermore, it would be very interesting to consider non-protected external operators as this would give us access to a vast number of correlators without increasing the number of operators in the OPE channel. It would  also be interesting to explore what would be the analogue of the integrated correlators constraints (\ref{eq:constr12}) in a non-protected setup, where integrability can produce a lot of information for the deformations away from the straight line with non-protected insertions at the cusp.

As we now managed to apply the \texttt{SDPB} optimisation script for non-standard asymptotics of the effective conformal blocks, one can try to apply some variation of this method to 
incorporate analytical sum-rules into the set of equations. Such relations emerge for example from Mellin space considerations \cite{Bianchi:2021piu}, the Polyakov bootstrap  \cite{Mazac:2018mdx,Mazac:2018ycv,Mazac:2018qmi,Paulos:2019fkw,Kaviraj:2022wbw}, or the study of bulk locality~\cite{Levine:2023ywq}, and could help to supplement (or extract more efficiently) the information coming from integrability.

Finally, it would be interesting to see how the knowledge of the spectrum plays together with the recent development in the 6-point bootstrap, see e.g. \cite{Bercini:2020msp,Buric:2020dyz,Antunes:2021kmm,Buric:2021ywo,Kaviraj:2022wbw}, which offers two additional continuous parameters into the crossing equation. This would require deriving the 6-point conformal blocks in the current SUSY setup. There are also analytical predictions for a 6-point function in this theory at weak~\cite{Barrat:2021tpn} and strong coupling~\cite{Giombi:2023zte}. Equally, it would be interesting to find bounds for the 6-point function as well.

Ultimately, one can also try to get narrow bounds by extending the current technology to the local operators (important progress in this direction was made in \cite{Caron-Huot:2022sdy}). For this problem, the spectrum is readily available from \cite{Marboe:2018ugv,Gromov:2023hzc,Julius:2023hre}, although obviously one needs to deal with spin and double trace contributions. 
Going further, it would be interesting to consider mixed bootstrap problems with operators on and off the defect as in the general setup of \cite{Billo:2016cpy} and recent works \cite{Barrat:2021yvp,Bianchi:2022ppi,Barrat:2022psm,Barrat:2020vch,Pufu:2023vwo,Billo:2023ncz}.

\acknowledgments
We are grateful to Ant\'onio Antunes, Lorenzo Bianchi, Gabriel Bliard, Simon Ekhammar, Pietro Ferrero, Valentina Forini, Johan Henriksson, Shota Komatsu, Petr Kravchuk, Nat Levine, Marco Meineri, Carlo Meneghelli, Miguel Paulos, Giulia Peveri, Paul Ryan, Bharath Radhakrishnan, Roberto Tateo and Emilio Trevisani for discussions, and especially to Julius Julius and Nika Sokolova for collaboration on closely related projects and helping us to test the integrability description for the parity charge using results from an upcoming paper~\cite{NikaJuliusFuture}. 
AC is supported by the INFN SFT specific initiative. The work of N.G. and a part of the work of M.P was
supported by the European Research Council (ERC) under the European Union’s Horizon 2020 research and innovation program – 60 – (grant agreement No. 865075) EXACTC. The work of MP is supported by Marie Sk\l{}odowska-Curie Global Fellowship (HORIZON-MSCA-2022-PF-01) BOOTSTRABILITY-101109934.

\appendix

\section{Superconformal Block Expansions}\label{app:app1}
The OPE decomposition of the correlator involves the supermultiplets
\cite{Liendo:2018ukf}
\beq\label{eq:OPEfusion}
\mathcal{B}_1 \times \mathcal{B}_1  = \mathbb{I} + \mathcal{B}_2 + \sum_{\Delta>1} \mathcal{L}_{0,[0,0]}^{\Delta} \;,
\eeq
where $\mathbb{I}$ denotes the identity block and $\mathcal{B}_2$ is a BPS multiplet.

The superconformal blocks entering the decomposition (\ref{eq:reducedOPE}) of the reduced correlator are given explicitly by
\beqa\label{superblocks}
F_{\mathbb{I}}(x) &=& x\;,\\
F_{\mathcal{B}_2}(x) &=& x - x\, _2F_1(1,2,4;x ) \;,\\
F_{{\Delta}}(x) &=& \frac{x^{\Delta+1}}{1-\Delta}\, _2F_1(\Delta+1,\Delta+2,2 \Delta+4;x )\; .
\eeqa

\section{Details on Integrated Correlators}\label{app:app2}
The integrated correlator constraints (\ref{eq:constr12}) read
\begin{align}
\int_{0}^{1/2}
     \left(f(x) - x +\frac{C^2_{{\rm BPS}}}{2} x^2\right)\mu_a(x) dx
     +\mathcal{K}_a(g) = 0\;\;,\;\;a=1,2\; ,\label{eq:constr12app}
\end{align}
where
the integration measures  are
\beqa
\mu_1(x) &=& \frac{1}{x-1}+\frac{(x-1)^2}{x^3}, \\
\mu_2(x) &=& \frac{2 x - 1}{x^2} .
\eeqa
The constants $\mathcal{K}_a$, $a = 1,2$ appearing in the identities (\ref{eq:constr12}) are
\beqa
\mathcal{K}_1(g) &=&  \frac{\mathbb{B} - 3 \mathbb{C} }{8 \mathbb{B}^2} + (1  -\mathbb{F}  ) \left(\frac{7}{8}  - \log(2) \right) + \log(2),\\
\mathcal{K}_2(g) &=& -\frac{\mathbb{C} }{4 \mathbb{B}^2 } + \frac{7}{8} (1- \mathbb{F} ) + 1 + \log(2) .
\eeqa
Putting these together with the contributions of the $\mathcal{B}_2$ block, in (\ref{intcorr2}) we get
\beqa\label{rhs}
\texttt{RHS}_1 &\equiv& \frac{\mathbb{B} - 3 \mathbb{C} }{8 \mathbb{B}^2} + (\mathbb{F} - 1) \left( 7\log(2) - \frac{41}{8} \right) + \log(2) , \\
\texttt{RHS}_2 &\equiv& - \frac{\mathbb{C} }{4 \mathbb{B}^2} + (1-\mathbb{F}) \left(\log(2) + \frac{1}{6} \right)+ 1 + \log(2). 
\eeqa
The quantities appearing in \eqref{rhs} are the Bremsstrahlung function $\mathbb{B}$ \cite{Correa:2012at,Fiol:2012sg,Erickson:2000af,Drukker:2000rr,Drukker:2006ga,Pestun:2009nn,Gromov:2012eu,Gromov:2013qga,Sizov:2013joa,Bonini:2015fng} and Curvature function $\mathbb{C}$ \cite{Gromov:2015dfa} which appear in the cusp anomalous dimension
\beq
\Gamma^{\text{cusp}}(\theta) = \mathbb{B}(g)\sin^2\theta  + \frac{1}{4} (\mathbb{B}(g) + \mathbb{C}(g) ) \sin^4\theta +O(\sin^6\theta )
\eeq
where $\theta$ parametrise the deviation in R-symmetry space from the straight Wilson line limit. 
The exact expression of $\mathbb{B}(g)$ and $\mathbb{C}(g)$ is the following 
\begin{align}
   \label{bremdef}\mathbb{B}(g) &= \frac{g}{\pi}\frac{I_2(4 \pi g)}{I_1(4 \pi g)}\;,\\
    \label{curvaturedef}\mathbb{C}(g) &= -4\,\mathbb{B}^2(g)
	-\frac{1}{2}\oint\frac{du_x}{2\pi i}\oint\frac{d
 u_y}{2\pi i}
    K_0(u_x-u_y)F[x, y]\;,
\end{align}
where $I_n$ are modified Bessel function of the first kind.
In the second line, both integrals run clockwise around the cut $[-2g,2g]$ with $u_x = g ( x + 1/x)$ the Zhukovsky parametrisation. The kernel $K_0$ and the integrand $F$ are given in the appendix of \cite{Cavaglia:2022qpg}. 
The quantity $\mathbb{F} $ is related to the OPE coefficient $C_{\rm BPS}^2$, and is given explicitly by~\cite{Liendo:2018ukf}
\beq
\mathbb{F}(g) = 1 + C^2_{\rm BPS}(g) = \frac{3 I_1(4 g \pi ) \left(\left(2 \pi ^2 g^2+1\right) I_1(4 g \pi )-2 g \pi  I_0(4 g \pi )\right)}{2 g^2 \pi ^2 I_2(4 g \pi ){}^2}.
\eeq

Integrating the two measures against the conformal blocks we get the following exact integrals
\begin{equation}\small\begin{split}\label{int1app}
\int_0^{1/2}
\mu_1(x)F_{\Delta}(x)dx=&\frac{-2^{-\Delta -1}3 \Delta  (\Delta  (\Delta +3)+4)}{\Delta  (\Delta +2) (\Delta +4) \left(\Delta ^2-1\right)^2} \biggl[\,_2F_1\left(\Delta +1,\Delta +3;2 (\Delta +2);\frac{1}{2}\right)\\
&\quad+\frac{(8+\Delta  (\Delta +3) (\Delta  (\Delta +3)+4))}{\Delta  (\Delta  (\Delta +3)+4)} \, _2F_1\left(\Delta ,\Delta +3;2 (\Delta +2);\frac{1}{2}\right)\biggl]
\end{split}\normalsize\end{equation}
and
\begin{equation}\small\begin{split}\label{int2app}
\int_0^{1/2}
\mu_2(x)
F_{\Delta}(x)dx=\frac{2^{-\Delta }}{\Delta  \left(\Delta ^2-1\right)} \biggl[(\Delta &+1) \, _2F_1\left(\Delta ,\Delta +2;2 (\Delta +2);\frac{1}{2}\right) ,\\
&-\Delta  \, _2F_1\left(\Delta +1,\Delta +1;2 (\Delta +2);\frac{1}{2}\right)\biggl]
\end{split}\normalsize.
\end{equation}
These functions (or their polynomial-type approximations) enter the last two entries of the vector $V_{\Delta}$ (cf., section \ref{sec:integratedbootstrap}) in the bootstrap algorithm. 
The integrals over the BPS block are given by
\begin{equation}\small\begin{split}\label{int1B2app}
\int_0^{1/2}
\mu_1(x)\left(F_{\mathcal{B}_2}(x)+\frac{x^2}{2}\right)dx=6\log 2-\frac{17}{4}\,,
\end{split}\normalsize\end{equation}
and 
\begin{equation}\small\begin{split}\label{int2B2app}
\int_0^{1/2}
\mu_2(x)\left(F_{\mathcal{B}_2}(x)+\frac{x^2}{2}\right)dx=-\log 2-\frac{17}{24}\,.
\end{split}\normalsize\end{equation}

\section{Bounds}\label{app:app3}
\subsection{OPE coefficients}
The data obtained for the OPE coefficients exploiting the parity symmetry are listed in Tables \ref{tab:C1bounds}-\ref{tab:C3bounds}. The format is $\frac{1}{2}\left(C^2_{i\text{ lower}} + C_{i\text{ upper}}^2\right) \pm \frac{1}{2}\left(C^2_{i\text{ upper}} - C_{i\text{ lower}}^2\right)$. The results are obtained with the input from the spectrum of the first 10 states excluding $\Delta_7$ and exploiting the two integrated correlator constraints. We use $\Lambda = 140$ and $\Lambda = 60$ with $N_{\text{poles}} = 30$. These results can be extracted from the \texttt{Mathematica} notebook attached to this paper where we include also a set of data for $C_1^2$ computed with $\Lambda = 60$.

\begin{table}[h!]
   \centering
    \begin{tabular}{||cc|cc||}
    \hline
    $g$ & $C_1^2$ & $g$ & $C_1^2$  \\
    \hline\hline
 0.1 & 0.0189694960
\,$\pm$\,4.26
 \,$10^{-8}$ & 2.0 & 0.34478716196
 \,$\pm$\,   1.56
 \, $10^{-9}$ \\
 0.2 & 0.065679029
 \,$\pm$\,6.95
 $10^{-7}$ & 2.2 & 0.34963125354
 \,$\pm$\,  1.02
 \, $ 10^{-9}$ \\
 0.4 & 0.16838882
 $\pm  $ 1.29
 \,$10^{-6}$ & 2.4 & 0.353696925683
 \,$\pm$\, 6.99
 \,$10^{-10}$ \\
 0.6 &  0.233041778
 \,$\pm$\, 4.01
 \,$10^{-7}$ & 2.6 & 0.357157434539
 \,$\pm$\, 4.94
 \,$10^{-10}$ \\
 0.8 &  0.270286755
 \,$\pm$\, 1.12
 \,$10^{-7}$ & 2.8 & 0.360138240800
 \,$\pm$\, 3.59
 \,$10^{-10}$\\
 1.0 & 0.2940148837
 \,$\pm$\, 3.88
 \,$10^{-8}$ & 3.0 & 0.362732415470
 \,$\pm$\, 2.67
 \,$10^{-10}$ \\
 1.2 & 0.3104333131
 \,$\pm$\, 1.63
 \,$10^{-8}$ & 3.2 & 0.365010449615
 \,$\pm$\, 2.03
 \,$10^{-10}$ \\
 1.4 & 0.32246686675
 \,$\pm$\, 7.96
 \,$10^{-9}$ & 3.4 & 0.367026704120
 \,$\pm$\, 1.58
 \,$10^{-10}$ \\
 1.6 & 0.33166329330
 \,$\pm$\, 4.29
 \,$10^{-9}$ & 3.6 & 0.368823769371
 \,$\pm$\, 1.24
 \,$10^{-10}$ \\
 1.8 & 0.33891847985
 \,$\pm$\, 2.51
 \,$10^{-9}$ & 3.8 & 0.3704354843206
\,$\pm$\, 9.99
\,$10^{-11}$ \\
    \hline
\end{tabular}
\caption{Bounds for the OPE coefficient $C_1^2$ at $\Lambda=140$.}
    \label{tab:C1bounds}
\end{table}

\begin{table}[h!]
   \centering
    \begin{tabular}{||cc|cc||}
    \hline
    $g$ & $C_2^2$ & $g$ & $C_2^2$  \\
    \hline\hline
 0.1 & 0.1008
\,$\pm$\,1.77
 \,$10^{-2}$ & 2.0 & 0.031809
 \,$\pm$\,   1.28
 \, $10^{-4}$ \\
 0.2 & 0.09425
 \,$\pm$\,9.41
 $10^{-3}$ & 2.2 & 0.031145
 \,$\pm$\,  1.07
 \, $ 10^{-4}$ \\
 0.4 & 0.06962
 $\pm  $ 3.82
 \,$10^{-3}$ & 2.4 & 0.0305987
 \,$\pm$\, 9.28
 \,$10^{-5}$ \\
 0.6 &  0.05276
 \,$\pm$\, 2.10
 \,$10^{-3}$ & 2.6 & 0.0301406
 \,$\pm$\, 8.15
 \,$10^{-5}$ \\
 0.8 &  0.04439
 \,$\pm$\, 1.04
 \,$10^{-3}$ & 2.8 & 0.0297512
 \,$\pm$\, 7.28
 \,$10^{-5}$\\
 1.0 & 0.039823
 \,$\pm$\, 5.90
 \,$10^{-4}$ & 3.0 & 0.0294160
 \,$\pm$\, 6.59
 \,$10^{-5}$ \\
 1.2 & 0.036993
 \,$\pm$\, 3.79
 \,$10^{-4}$ & 3.2 & 0.0291243
 \,$\pm$\, 6.04
 \,$10^{-5}$ \\
 1.4 & 0.035073
 \,$\pm$\, 2.66
 \,$10^{-4}$ & 3.4 & 0.0288683
 \,$\pm$\, 5.59
 \,$10^{-5}$ \\
 1.6 & 0.033685
 \,$\pm$\, 1.99
 \,$10^{-4}$ & 3.6 & 0.0286416
 \,$\pm$\, 5.22
 \,$10^{-5}$ \\
 1.8 & 0.032634
 \,$\pm$\, 1.56
 \,$10^{-4}$ & 3.8 & 0.0284396
\,$\pm$\, 4.91
\,$10^{-5}$ \\
    \hline
\end{tabular}
\caption{Bounds for the OPE coefficient $C_2^2$ at $\Lambda=60$.}
    \label{tab:C2bounds}
\end{table}

\begin{table}[h!]
   \centering
    \begin{tabular}{||cc|cc||}
    \hline
    $g$ & $C_3^2$ & $g$ & $C_3^2$  \\
    \hline\hline
 0.1 & 0.1014
\,$\pm$\,2.11
 \,$10^{-2}$ & 2.0 & 0.137422
 \,$\pm$\,   1.42
 \, $10^{-4}$ \\
 0.2 & 0.1105
 \,$\pm$\,1.63
 $10^{-2}$ & 2.2 & 0.136093
 \,$\pm$\,  1.17
 \, $ 10^{-4}$ \\
 0.4 & 0.13103
 $\pm  $ 9.52
 \,$10^{-3}$ & 2.4 & 0.1349216
 \,$\pm$\, 9.99
 \,$10^{-5}$ \\
 0.6 &  0.14487
 \,$\pm$\, 4.19
 \,$10^{-3}$ & 2.6 & 0.1338840
 \,$\pm$\, 8.69
 \,$10^{-5}$ \\
 0.8 &  0.14780
 \,$\pm$\, 1.66
 \,$10^{-3}$ & 2.8 & 0.1329608
 \,$\pm$\, 7.70
 \,$10^{-5}$\\
 1.0 & 0.146711
 \,$\pm$\, 8.25
 \,$10^{-4}$ & 3.0 & 0.1321350
 \,$\pm$\, 6.92
 \,$10^{-5}$ \\
 1.2 & 0.144679
 \,$\pm$\, 4.86
 \,$10^{-4}$ & 3.2 & 0.1313928
 \,$\pm$\, 6.31
 \,$10^{-5}$ \\
 1.4 & 0.142582
 \,$\pm$\, 3.23
 \,$10^{-4}$ & 3.4 & 0.1307226
 \,$\pm$\, 5.81
 \,$10^{-5}$ \\
 1.6 & 0.140651
 \,$\pm$\, 2.32
 \,$10^{-4}$ & 3.6 & 0.1301147
 \,$\pm$\, 5.40
 \,$10^{-5}$ \\
 1.8 & 0.138934
 \,$\pm$\, 1.77
 \,$10^{-4}$ & 3.8 & 0.1295611
\,$\pm$\, 5.06
\,$10^{-5}$ \\
    \hline
\end{tabular}
\caption{Bounds for the OPE coefficient $C_3^2$ at $\Lambda=60$.}
    \label{tab:C3bounds}
\end{table}

\subsection{Four-point function}
The data obtained for four-point function $G$ and the reduced correlator $f$ are listed in Tables \ref{tab:GNBPSbounds} and \ref{tab:fNBPSbounds}. The format is the same used for the OPE coefficients above. The results are obtained with the input from the spectrum of the first 10 states excluding $\Delta_7$ and exploiting the two integrated correlator constraints. We use $\Lambda = 90$ with $N_{\text{poles}} = 40$ for all data. Also these results can be extracted from the \texttt{Mathematica} notebook attached to this paper were we include other set of data computed for different values of $\Lambda$ and $N_{\text{poles}}$.

\begin{table}[h!]
   \centering
    \begin{tabular}{||ccc|ccc||}
    \hline
    $g$ & $x$ & $G_\text{NBPS}/H_{\Delta_1}$ & $g$ & $x$ & $G_\text{NBPS}/H_{\Delta_1}$  \\
    \hline\hline
\multirow{5}{*}{0.02} & 0.02 & 0.003648663
\,$\pm$\,6.06
 \,$10^{-7}$ & \multirow{5}{*}{2.20} & 0.02 & 0.349821502794
 \,$\pm$\,   2.93
 \, $10^{-10}$ \\
  & 0.14 & 0.002197471
 \,$\pm$\,1.74
 $10^{-7}$ &  & 0.14 & 0.355756456661
 \,$\pm$\,  4.23
 \, $ 10^{-10}$ \\
  & 0.26 & 0.0017323312
 $\pm  $ 1.21
 \,$10^{-8}$ &  & 0.26 & 0.36889975844
 \,$\pm$\, 1.12
 \,$10^{-9}$ \\
  & 0.38 &  0.0016522652
 \,$\pm$\, 1.48
 \,$10^{-8}$ &  & 0.38 & 0.389577801438
 \,$\pm$\, 2.21
 \,$10^{-10}$ \\
  & 0.50 &  0.0017449097
 \,$\pm$\, 1.22
 \,$10^{-8}$ &  & 0.50 & 0.421024281568
 \,$\pm$\, 4.89
 \,$10^{-10}$\\[0.5em]
\multirow{5}{*}{0.40} & 0.02 & 0.17040454
\,$\pm$\,3.25
 \,$10^{-6}$ & \multirow{5}{*}{2.80} & 0.02 &  0.3602921472699
 \,$\pm$\,   4.69
 \, $10^{-11}$ \\
  & 0.14 &  0.18925256
 \,$\pm$\,3.50
 $10^{-6}$ &  & 0.14 &  0.365586079936
 \,$\pm$\,  1.08
 \, $ 10^{-10}$ \\
  & 0.26 &  0.21122069
 $\pm  $ 1.16
 \,$10^{-6}$ &  & 0.26 & 0.377850185762
 \,$\pm$\, 4.44
 \,$10^{-10}$ \\
  & 0.38 &   0.234136340
 \,$\pm$\, 9.01
 \,$10^{-7}$ &  & 0.38 & 0.3976737621335
 \,$\pm$\, 7.70
 \,$10^{-11}$ \\
  & 0.50 &  0.262704154
 \,$\pm$\, 9.82
 \,$10^{-7}$ &  & 0.50 &  0.428334782014
 \,$\pm$\, 1.88
 \,$10^{-10}$\\[0.5em]
 \multirow{5}{*}{1.00} & 0.02 & 0.2944930912
\,$\pm$\,3.77
 \,$10^{-8}$ & \multirow{5}{*}{3.40} & 0.02 & 0.36715948436860
 \,$\pm$\,   8.23
 \, $10^{-12}$ \\
  & 0.14 & 0.3041446748
 \,$\pm$\,3.52
 $10^{-8}$ &  & 0.14 & 0.3720468584919
 \,$\pm$\,  3.44
 \, $ 10^{-11}$ \\
  & 0.26 &  0.3215050072
 $\pm  $ 3.29
 \,$10^{-8}$ &  & 0.26 &  0.383725154158
 \,$\pm$\, 2.19
 \,$10^{-10}$ \\
  & 0.38 &  0.3455612192
 \,$\pm$\, 1.07
 \,$10^{-8}$ &  & 0.38 & 0.4029548348299
 \,$\pm$\, 3.45
 \,$10^{-11}$ \\
  & 0.50 &  0.3794751685
 \,$\pm$\, 1.69
 \,$10^{-8}$ &  & 0.50 & 0.4330471746390
 \,$\pm$\, 9.16
 \,$10^{-11}$\\[0.5em]
 \multirow{5}{*}{1.60} & 0.02 & 0.33192802082
\,$\pm$\,2.23
 \,$10^{-9}$ & \multirow{5}{*}{4.00} & 0.02 & 0.37200817194817
 \,$\pm$\,  8.87
 \, $ 10^{-12}$ \\
  & 0.14 &  0.33901054548
 \,$\pm$\,2.48
 $10^{-9}$ &  & 0.14 &  0.3766156376958
 \,$\pm$\,  1.18
 \, $ 10^{-11}$ \\
  & 0.26 & 0.35360649476
 $\pm  $ 4.15
 \,$10^{-9}$ &  & 0.26 &  0.387876766918
 \,$\pm$\, 1.24
 \,$10^{-10}$ \\
  & 0.38 &  0.375594057501
 \,$\pm$\, 9.83
 \,$10^{-10}$ &  & 0.38 & 0.4066717494335
 \,$\pm$\, 1.82
 \,$10^{-11}$ \\
  & 0.50 &   0.40815148693
 \,$\pm$\, 1.89
 \,$10^{-9}$ &  & 0.50 & 0.4363372525205
 \,$\pm$\, 5.17
 \,$10^{-11}$\\
    \hline
\end{tabular}
\caption{Bounds for the four-point function $G_\text{NBPS}/H_{\Delta_1}$.}
    \label{tab:GNBPSbounds}
\end{table}

\begin{table}[h!]
   \centering
    \begin{tabular}{||ccc|ccc||}
    \hline
    $g$ & $x$ & $f_\text{NBPS}/H_{\Delta_1}$ & $g$ & $x$ & $f_\text{NBPS}/H_{\Delta_1}$  \\
    \hline\hline
\multirow{5}{*}{0.02} & 0.02 & 0.00080475237
\,$\pm$\,5.35
 \,$10^{-9}$ & \multirow{5}{*}{2.20} & 0.02 & 0.349693009833
 \,$\pm$\,   7.86
 \, $10^{-10}$ \\
  & 0.10 & 0.00083315205
 \,$\pm$\,2.26
 $10^{-9}$ &  & 0.10 &  0.350722340900
 \,$\pm$\,  7.88
 \, $ 10^{-10}$ \\
  & 0.22 & 0.00088500607
 $\pm  $ 3.07
 \,$10^{-9}$ &  & 0.22 &  0.3545181908590
 \,$\pm$\, 1.31
 \,$10^{-11}$ \\
  & 0.34 &  0.00095282279
 \,$\pm$\, 3.16
 \,$10^{-9}$ &  & 0.34 &   0.361756700148
 \,$\pm$\, 3.31
 \,$10^{-10}$ \\
  & 0.46 &  0.001046222070
 \,$\pm$\, 8.29
 \,$10^{-10}$ &  & 0.46 & 0.3742588781538
 \,$\pm$\, 8.81
 \,$10^{-11}$\\[0.5em]
\multirow{5}{*}{0.40} & 0.02 &  0.168866552
\,$\pm$\,7.29
 \,$10^{-7}$ & \multirow{5}{*}{2.80} & 0.02 &  0.360188604141
 \,$\pm$\,   3.33
 \, $10^{-10}$ \\
  & 0.10 &  0.171713536
 \,$\pm$\,7.54
 $10^{-7}$ &  & 0.10 &  0.361099922095
 \,$\pm$\,  3.00
 \, $ 10^{-10}$ \\
  & 0.22 &  0.177876249
 $\pm  $ 3.85
 \,$10^{-7}$ &  & 0.22 & 0.3646175851682
 \,$\pm$\, 1.12
 \,$10^{-10}$ \\
  & 0.34 &   0.186709733
 \,$\pm$\, 4.58
 \,$10^{-7}$ &  & 0.34 & 0.371503342528
 \,$\pm$\, 1.729
 \,$10^{-10}$ \\
  & 0.46 &  0.199555629
 \,$\pm$\, 1.12
 \,$10^{-7}$ &  & 0.46 &  0.3835946548807
 \,$\pm$\, 3.48
 \,$10^{-11}$\\[0.5em]
 \multirow{5}{*}{1.00} & 0.02 & 0.2941609107
\,$\pm$\,1.07
 \,$10^{-8}$ & \multirow{5}{*}{3.40} & 0.02 & 0.367070355080
 \,$\pm$\,   1.70
 \, $10^{-10}$ \\
  & 0.10 &  0.2958445094
 \,$\pm$\,2.49
 $10^{-8}$ &  & 0.10 & 0.367906457825
 \,$\pm$\,  1.44
 \, $ 10^{-10}$ \\
  & 0.22 &  0.30092381206
 $\pm  $ 3.78
 \,$10^{-9}$ &  & 0.22 &  0.37123782985616
 \,$\pm$\, 9.02
 \,$10^{-12}$ \\
  & 0.34 &  0.3095818565
 \,$\pm$\, 1.05
 \,$10^{-8}$ &  & 0.34 & 0.3778792495534
 \,$\pm$\, 6.31
 \,$10^{-11}$ \\
  & 0.46 &  0.32352970244
 \,$\pm$\, 2.66
 \,$10^{-9}$ &  & 0.46 & 0.3896778779563
 \,$\pm$\, 1.72
 \,$10^{-11}$\\[0.5em]
 \multirow{5}{*}{1.60} & 0.02 &  0.33174781876
\,$\pm$\,2.46
 \,$10^{-9}$ & \multirow{5}{*}{4.00} & 0.02 & 0.3719283405376
 \,$\pm$\,  9.93
 \, $ 10^{-11}$ \\
  & 0.10 &  0.33298555343
 \,$\pm$\,3.03
 $10^{-9}$ &  & 0.10 &  0.3727125678220
 \,$\pm$\,  8.01
 \, $ 10^{-11}$ \\
  & 0.22 & 0.337235152044
 $\pm  $ 1.63
 \,$10^{-10}$ &  & 0.22 &  0.37591099865378
 \,$\pm$\, 6.68
 \,$10^{-12}$ \\
  & 0.34 &   0.34501726431
 \,$\pm$\, 1.25
 \,$10^{-9}$ &  & 0.34 & 0.3823740429946
 \,$\pm$\, 3.57
 \,$10^{-11}$ \\
  & 0.46 &   0.358121053143
 \,$\pm$\, 3.26
 \,$10^{-10}$ &  & 0.46 & 0.39395510243309
 \,$\pm$\, 9.82
 \,$10^{-12}$\\
    \hline
\end{tabular}
\caption{Bounds for the four-point function $f_\text{NBPS}/H_{\Delta_1}$.}
    \label{tab:fNBPSbounds}
\end{table}
\newpage 
\bibliographystyle{JHEP.bst}
\bibliography{references}

\end{document}